\documentclass[10pt, twocolumn, twoside]{IEEEtran}

\usepackage{amsmath}
\usepackage{epsfig,amssymb,amsfonts}
\usepackage{amsbsy,bbm}
\usepackage{stmaryrd}
\usepackage[latin1]{inputenc}
\usepackage{verbatim}
\usepackage{graphicx}
\usepackage{graphics}
\usepackage{theorem}
\usepackage{tabularx}
\usepackage{array}
\usepackage{multicol,float}
\usepackage[colorlinks=true, pdfstartview=FitV, linkcolor=blue, citecolor=blue, urlcolor=blue]{hyperref}
\usepackage{hhline}
 \usepackage{epstopdf}
\usepackage[T1]{fontenc}
\usepackage{tikz}
\usepackage{tkz-berge}
\usepackage{pgfplots}
\usetikzlibrary{plotmarks}
\usepackage{subfigure}
\usepackage{cite}

\usepackage[babel=true]{csquotes}

\usepackage{times}

\usepackage{color} 

\definecolor{lightgray}{gray}{0.5}
\definecolor{gris}{rgb}{0.6,0.6,0.6} 
\definecolor{matlab}{rgb}{0,0.6,0} 

\usepackage{graphicx} 
\usepackage{graphicx,subfigure}
\usepackage{amssymb,amsmath,amsfonts} 
\usepackage{bm} 
\usepackage{bbm} 
\usepackage{dsfont}
\usepackage[overload]{empheq} 
\usepackage{cases} 
\usepackage{mathrsfs} 
\newcommand{\R}{\mathbb{R}}
\newcommand{\C}{\mathbb{C}}
\newcommand{\Z}{\mathbb{Z}}
\newcommand{\N}{\mathbb{N}}
\newcommand{\E}{\mathbb{E}}
\newcommand{\Nul}{\mathbbm{O}}
\newcommand{\NC}{\mathcal{CN}}

\title{Convergence of Structured Quadratic Forms With Application to Theoretical Performances of Adaptive Filters in Low Rank Gaussian Context}
\author{Alice~Combernoux,~\IEEEmembership{Student Member,~IEEE}, Fr\'{e}d\'{e}ric~Pascal,~\IEEEmembership{Senior Member,~IEEE}, Guillaume~Ginolhac,~\IEEEmembership{Member,~IEEE}~and~Marc~Lesturgie~\IEEEmembership{Senior Member,~IEEE}
\thanks{Alice Combernoux and Marc Lesturgie are with SONDRA - ONERA, CentraleSupelec, Fr\'{e}d\'{e}ric Pascal is with L2S - CentraleSupelec - CNRS - Universit\'{e} Paris-Sud 11, Guillaume Ginolhac is with LISTIC - Universit\'{e} Savoie Mont Blanc}}

\begin{document}

\maketitle

\begin{abstract}
This paper addresses the problem of deriving the asymptotic performance of adaptive Low Rank (LR) filters used in target detection embedded in a disturbance composed of a LR Gaussian noise plus a white Gaussian noise. In this context, we use the Signal to Interference to Noise Ratio (SINR) loss as performance measure which is a function of the estimated projector onto the LR noise subspace. However, although the SINR loss can be determined through Monte-Carlo simulations or real data, this process remains quite time consuming. Thus, this paper proposes to predict the SINR loss behavior in order to not depend on the data anymore and be quicker. To derive this theoretical result, previous works used a restrictive hypothesis assuming that the target is orthogonal to the LR noise. In this paper, we propose to derive this theoretical performance by relaxing this hypothesis and using Random Matrix Theory (RMT) tools. These tools will be used to present the convergences of \textit{simple} quadratic forms and perform new RMT convergences of \textit{structured} quadratic forms and SINR loss in the large dimensional regime, i.e. the size and the number of the data tend to infinity at the same rate. We show through simulations the interest of our approach compared to the previous works when the restrictive hypothesis is no longer verified.

\end{abstract}


\section{Introduction}
In array processing, the covariance matrix $\mathbf{R}$ of the data is widely involved for main applications as filtering~\cite{ReMaBr74,Wa94}, radar/sonar detection~\cite{ScFr94} or localization~\cite{Sc86,RoKa89}. However, when the disturbance in the data is composed of the sum of a Low Rank~(LR) correlated noise and a White Gaussian Noise~(WGN), the covariance matrix is often replaced by the projector onto the LR noise subspace $\mathbf{\Pi}_{\mathrm{c}}$~\cite{KiTu94,Ha96,GiJo02,RaLiGe04}. In practice, the projector onto the LR noise subspace (and the covariance matrix) is generally unknown and an estimate is consequently required to perform the different processing. This estimation procedure is based on the so-called secondary data assumed to be independent and to share the same distribution. Then, the true projector is replaced by the estimated one in order to obtain an adaptive processing. An important issue is then to derive the theoretical performances of the adaptive processing as a function of the number of secondary data $K$. The processing based on the covariance matrix has been widely studied and led to many theoretical results in filtering~\cite{ReMaBr74} and detection~\cite{Ke86,RoFuKeNi92,KrScMc01,BeSc06}. For example, for classical adaptive processing, $K=2m$ secondary data (where $m$ is the data size) are required to ensure good performance of the adaptive filtering, i.e. a 3dB loss of the output Signal to Interference plus Noise Ratio (SINR) compared to optimal filtering~\cite{ReMaBr74}. For LR processing, some results has been obtained especially in filtering~\cite{KiTu94,Ha97,PeHaAyGoRe00,GiFo13} and localization~\cite{KrFoPr92}. Similarly, in LR filtering, the number $K$ of secondary data required to ensure good performance of the adaptive filtering is equal to $2r$ (where $r$ is the rank of the LR noise subspace)~\cite{KiTu94,Ha97}.

These last results are obtained from the theoretical study of the Signal to SINR loss. More precisely, in~\cite{Ha97,GiFo13}, the derivation of the theoretical results is based on the hypothesis that the steering vector is orthogonal to the LR noise subspace. Nevertheless, even if the result seems to be close to the simulated one when the hypothesis is no longer valid anymore~\cite{GiFoPaOv14}, it is impossible with traditional techniques of~\cite{Ha97,GiFo13} to obtain a theoretical performance as a function of the distance between the steering vector and the LR noise subspace. Since, in practice, this dependence is essential to predict the performance of the adaptive filtering, we propose in this paper to derive the theoretical SINR loss, for a disturbance composed of a LR noise and a WGN, as a function of $K$ and the distance between the steering vector and the LR noise subspace. The proposed approach is based on the study of the SINR loss structure.

The SINR loss (resp. LR SINR loss) is composed of a \textit{simple} Quadratic Form (QF) in the numerator, $\bm{s}_1^H \hat{\mathbf{R}}^{-1}\bm{s}_2$ (resp. $\bm{s}_1^H \hat{\mathbf{\Pi}}^{\bot}_{\mathrm{c}}\bm{s}_2$) and a \textit{structured} QF in the denominator $\bm{s}_1^H \hat{\mathbf{R}}^{-1}\mathbf{R}\hat{\mathbf{R}}^{-1}\bm{s}_2$ (resp. $\bm{s}_1^H \hat{\mathbf{\Pi}}^{\bot}_{\mathrm{c}}\mathbf{R}\hat{\mathbf{\Pi}}^{\bot}_{\mathrm{c}}\bm{s}_2$). These recent years, the \textit{simple} QFs (numerator) have been broadly studied~\cite{Me08,Me08bis,VaLoMe12,CoHa13} using Random Matrix Theory (RMT) tools contrary to \textit{structured} QFs (denominator). RMT tools have also been used in array processing to improve the MUSIC algorithm~\cite{MeLa08,CoPaSi14} and in matched subspace detection~\cite{NaSi10,AsNa13} where the rank $r$ is unknown. The principle is to examine the spectral behavior of $\mathbf{\hat{R}}$ by RMT to obtain their convergences, performances and asymptotic distribution when $K$ tends to infinity and when both the data size $m$ and $K$ tend to infinity at the same ratio, i.e. $m/K\rightarrow c\in ]0,+\infty)$, for different models of $\mathbf{\hat{R}}$ of the observed data as in~\cite{Me08,Me08bis,MeLa08},~\cite{CoHa13} and~\cite{VaLoMe12}. Therefore, inspired by these works, we propose in this paper to study the convergences of the \textit{structured} QFs $\bm{s}_1^H \hat{\mathbf{R}}^{-1}\mathbf{R}\hat{\mathbf{R}}^{-1}\bm{s}_2$ (resp. $\bm{s}_1^H \hat{\mathbf{\Pi}}^{\bot}_{\mathrm{c}}\mathbf{R}\hat{\mathbf{\Pi}}^{\bot}_{\mathrm{c}}\bm{s}_2$): when 1) $K\rightarrow\infty$ with a fixed $m$ and when 2) $m,K\rightarrow\infty$ at the same ratio under the most appropriated model for our data and with the rank assumed to be known. From~\cite{CoPaGiLe14,CoPaGiLe15}, the \textit{spiked} model has proved to be the more appropriated one to our knowledge. This model, introduced by~\cite{Jo01} (also studied in~\cite{BeNa11,Pa07} from an eigenvector point of view) considers that the multiplicity  of the eigenvalues corresponding to the signal (the LR noise for us) is fixed for all $m$ and leads to the SPIKE-MUSIC estimator~\cite{HaLoMeNaVa13} of $\bm{s}_1^H\bm{\hat{\Pi}}\bm{s}_2$. Then, the new results are validated through numerical simulations. From these new theoretical convergences, the paper derives the convergence of the SINR loss for both adaptive filters (the classical and the LR one). The new theoretical SINR losses depend on the number of secondary data $K$ but also on the distance between the steering vector and the LR noise subspace. This work is partially related to those of~\cite{TaTaPe10,TaTaPe13} and \cite{YuRuMc13} which uses the RMT tools to derive the theoretical SINR loss in a full rank context (previously defined as classical). 

Finally, these theoretical SINR losses are validated in a jamming application context where the purpose is to detect a target thanks to a Uniform Linear Antenna (ULA) composed of $m$ sensors despite the presence of jamming. The response of the jamming is composed of signals similar to the target response. This problem is very similar to the well-known Space Time Adaptive Processing (STAP) introduced in~\cite{Wa94}. Results show the interest of our approach with respect to other theoretical results~\cite{KiTu94,Ha97,PeHaAyGoRe00,GiFo13} in particular when the target is close to the jammer.

The paper is organized as follows. Section~\ref{sec:pb_statement} presents the received data model, the adaptive filters and the corresponding SINR losses. Section~\ref{sec:RMT} summarizes the existing studies on the \textit{simple} QFs $\bm{s}_1^H\mathbf{\hat{R}}\bm{s}_2$ and $\bm{s}_1^H\mathbf{\hat{\Pi}}\bm{s}_2$, and exposes the covariance matrix model, the \textit{spiked} model. Section~\ref{sec:NewCVResults} gives the theoretical contribution the paper with the convergences of the \textit{structured} QFs $\bm{s}_1^H\mathbf{\hat{\Pi}}_{\mathrm{c}}^\bot\mathbf{B}\mathbf{\hat{\Pi}}_{\mathrm{c}}^\bot\bm{s}_2$ and $\bm{s}_1^H\mathbf{\hat{\Pi}}_{\mathrm{c}}^\bot\mathbf{R}\mathbf{\hat{\Pi}}_{\mathrm{c}}^\bot\bm{s}_2$ and the convergences of the SINR losses. The results are finally applied on a jamming application in Section~\ref{sec:simu}.\\
\indent \textit{Notations:} The following conventions are adopted. An italic letter stands for a scalar quantity, boldface lowercase (uppercase) characters stand for vectors (matrices) and $(.)^H$ stands for the conjugate transpose. $\mathbf{I}_{N}$ is the $N\times N$ identity matrix, $\mathrm{tr}(.)$ denotes the trace operator and $\mathrm{diag}(.)$ denotes the diagonalization operator such as $(\mathbf{A})_{i,i}=(\mathrm{diag}(\mathbf{a}))_{i,i}=(\mathbf{a})_{i}$ and $(\mathbf{A})_{i,j}=0$ if $i\neq j$. $\#\left\lbrace\mathcal{A}\right\rbrace $ denotes the cardinality of the set $\mathcal{A}$. $[\![a,b]\!]$ is the set defined by $\left\lbrace x\in\Z:a\leqslant x\leqslant b,\forall(a,b)\in\Z^2\right\rbrace$. $\boldsymbol{\Nul}_{n\times N}$ is a $n\times N$ matrix full of 0. The abbreviations iid and a.s. stem for \textit{independent and identically distributed} and almost surely respectively.


\section{Problem statement}\label{sec:pb_statement}
\indent The aim of the problem is to filter the received observation vector $\boldsymbol{x}\in\C^{m\times 1}$ in order to whiten the noise without mitigating an eventual complex signal of interest $\boldsymbol{d}$ (typically a target in radar processing). In this paper, $\boldsymbol{d}$ will be a target response and is equal to $\alpha\boldsymbol{a}(\bm{\Theta})$ where $\alpha$ is an unknown complex deterministic parameter (generally corresponding to the target amplitude), $\boldsymbol{a}(\bm{\Theta})$ is the steering vector and $\bm{\Theta}$ is an unknown deterministic vector containing the different parameters of the target (e.g. the localization, the velocity, the Angle of Arrival (AoA), etc.). In the remainder of the article, in order to simplify the notations, $\bm{\Theta}$ will be omitted of the steering vector which will simply be denoted as $\boldsymbol{a}$. If necessary, the original notation will be taken.\\
\indent This section will first introduces the data model. Then, the filters, adaptive filters and the corresponding SINR loss, the quantity characterizing their performances, will be defined.
\subsection{Data model}\label{subsec:data_model}
\indent The observation vector can be written as:
\begin{eqnarray}
  \boldsymbol{x}=\boldsymbol{d}+\boldsymbol{c}+\boldsymbol{b}
  \label{eq:probdetectLR}
\end{eqnarray}
\noindent where $\boldsymbol{c}+\boldsymbol{b}$ is the noise that has to be whitened. $\boldsymbol{b}\in\C^{m\times 1}\sim\mathcal{CN}(\mathbf{0},\sigma^2\mathbf{I}_m)$ is an Additive WGN (AWGN) and $\boldsymbol{c}$ is a LR Gaussian noise $\boldsymbol{c}\in\C^{m\times 1}$ modeled by a zero-mean complex Gaussian vector with a normalized covariance matrix $\mathbf{C}$ ($\mathrm{tr}(\mathbf{C}) = m$), i.e. $\boldsymbol{c}\sim \mathcal{CN}(\mathbf{0},\mathbf{C})$. Consequently, the covariance matrix of the noise $\boldsymbol{c}+\boldsymbol{b}$ can be written as $\mathbf{R}=\mathbf{C}+\sigma^2\mathbf{I}_m\in\C^{m\times m}$. Moreover, considering a LR Gaussian noise, one has $\mathrm{rank}\left( \mathbf{C} \right) = r \ll m $ and hence, the eigendecomposition of $\mathbf{C}$ is:
\begin{eqnarray}
  \mathbf{C} = \sum_{i=1}^r \gamma_i \boldsymbol{u}_i\boldsymbol{u}_i^H
  \label{eq:SVDC}
\end{eqnarray}
where $\gamma_i$ and $\boldsymbol{u}_i$, $i\in[\![1;r]\!]$ are respectively the non-zero eigenvalues and the associated eigenvectors of $\mathbf{C}$, unknown in practice. This leads to:
\begin{eqnarray}
  \mathbf{R}=\sum_{i=1}^m\lambda_i\boldsymbol{u}_i\boldsymbol{u}_i^{H}
  \label{eq:R}
\end{eqnarray}
\noindent where $\lambda_i$ and $\boldsymbol{u}_i$, $i\in[\![1,m]\!]$ are respectively the eigenvalues and the associated eigenvectors of $\mathbf{R}$ with $\lambda_1 = \gamma_1+\sigma^2>\cdots>\lambda_r=\gamma_r+\sigma^2>\lambda_{r+1}=\cdots=\lambda_m=\sigma^2$. Then, the projector onto the LR Gaussian noise subspace $\boldsymbol{\Pi}_\mathrm{c}$ and the projector onto the orthogonal subspace to the LR Gaussian noise subspace $\boldsymbol{\Pi}_\mathrm{c}^{\bot}$ are defined as follows:
\begin{eqnarray}
	\begin{cases}
		\boldsymbol{\Pi}_\mathrm{c} = \sum_{i=1}^r\boldsymbol{u}_i\boldsymbol{u}_i^{H}\\
		\boldsymbol{\Pi}_\mathrm{c}^{\bot} =  \mathbf{I}_{m} -  \boldsymbol{\Pi}_\mathrm{c}=\sum_{i=r+1}^{m} {\boldsymbol{u}_i\boldsymbol{u}_i^{H}} 
	\end{cases}\label{eq:defprojectors}
\end{eqnarray}
\indent However, in practice, the covariance matrix $\mathbf{R}$ of the noise is unknown. Consequently, it is traditionally estimated with the Sample Covariance Matrix (SCM) which is computed from $K$ iid secondary data $\boldsymbol{x}_k=\boldsymbol{c}_k+\boldsymbol{b}_k$, $k\in[\![1,K]\!]$, and can be written as:
\begin{eqnarray}
  \hat{\mathbf{R}}=\frac{1}{K} \sum_{k=1}^{K} \boldsymbol{x}_k \boldsymbol{x}_k^{H}= \sum_{i=1}^m  \hat{\lambda}_i \hat{\boldsymbol{u}}_i \hat{\boldsymbol{u}}_i^H
  \label{eq:Rscm}
\end{eqnarray}
\noindent where $\hat{\lambda}_i$ and $\hat{\boldsymbol{u}}_i$, $i\in[\![1,m]\!]$ are respectively the eigenvalues and the eigenvectors of $\hat{\mathbf{R}}$ with $\hat{\lambda}_1 \geqslant\hat{\lambda}_2\geqslant\cdots\geqslant\hat{\lambda}_m$, $\boldsymbol{c}_k\sim\mathcal{CN}(\mathbf{0},\mathbf{C})$ and $\boldsymbol{b}_k\sim\mathcal{CN}(\mathbf{0},\sigma^2\mathbf{I}_m)$. Finally, the traditional estimated projectors estimated from the SCM are:
\begin{eqnarray}
  \begin{cases}
    \hat{\boldsymbol{\Pi}}_{\mathrm{c}}=\sum_{i=1}^r {\hat{\boldsymbol{u}}_i\hat{\boldsymbol{u}}_i^{H}} \\
    \hat{\boldsymbol{\Pi}}_{\mathrm{c}}^{\bot} =\mathrm{\textbf{I}}_{m} -  \hat{\boldsymbol{\Pi}}_{\mathrm{c}}= \sum_{i=r+1}^{m} {\hat{\boldsymbol{u}}_i\hat{\boldsymbol{u}}_i^{H}}, 
  \end{cases}
  \label{eq:PIcSCM}
\end{eqnarray}
\subsection{Adaptive filters}
\indent A filtering preprocessing on the observation vector $\boldsymbol{x}$ (Eq.(\ref{eq:probdetectLR})) is first done with the filter $\boldsymbol{w}$ in order to whiten the received signal $p=\boldsymbol{w}^H\boldsymbol{x}$. The filter maximizing the SINR is given by:
\begin{eqnarray}
  \boldsymbol{w}_\mathrm{opt}=\mathbf{R}^{-1}\boldsymbol{a}
  \label{eq:wopt}
\end{eqnarray}
\noindent Since, in practice, the covariance matrix $\mathbf{R}$ of the noise is unknown, the estimated optimal filter or adaptive filter (sub-optimal) is:
\begin{eqnarray}
  \boldsymbol{\hat{w}}=\mathbf{\hat{R}}^{-1}\boldsymbol{a}
  \label{eq:wSCM}
\end{eqnarray}
\indent In the case where one would benefit of the LR structure of the noise, one should use the optimal LR filter, based on the fact that $\boldsymbol{\Pi}_\mathrm{c}^{\bot}$ is the best rank $r$ approximation of $\mathbf{R}^{-1}$, which is defined by~\cite{KiTu94}:
\begin{eqnarray}
  \boldsymbol{w}_\mathrm{LR}=\boldsymbol{\Pi}_\mathrm{c}^{\bot}\boldsymbol{a}
  \label{eq:wLRopt}
\end{eqnarray}
\noindent As, in practice, the projector is not known and is estimated from the SCM, the estimated optimal filter or adaptive filter (sub-optimal) is:
\begin{eqnarray}
  \boldsymbol{\hat{w}}_\mathrm{LR}=\hat{\boldsymbol{\Pi}}_{\mathrm{c}}^{\bot}\boldsymbol{a}
  \label{eq:wLRSCM}
\end{eqnarray}
\subsection{SINR Loss}
Then, we define the SINR Loss. In order to characterize the performance of the estimated filters, the SINR loss compares the SINR at the output of the filter to the maximum SINR: \small
\begin{eqnarray}
  \hat{\rho}&=&\frac{SINR_{out}}{SINR_{max}}=\frac{\vert\boldsymbol{\hat{w}}^H\boldsymbol{d}\vert^2}{(\boldsymbol{\hat{w}}^H\mathbf{R}\boldsymbol{\hat{w}})(\boldsymbol{d}^H\mathbf{R}^{-1}\boldsymbol{d})}\\
  &=&\frac{\vert\boldsymbol{a}^H\mathbf{\hat{R}}^{-1}\boldsymbol{a}\vert^2}{(\boldsymbol{a}^H\mathbf{\hat{R}}^{-1}\mathbf{R}\mathbf{\hat{R}}^{-1}\boldsymbol{a})(\boldsymbol{a}^H\mathbf{R}^{-1}\boldsymbol{a})}
  \label{eq:SNRLoss_wSCM}
\end{eqnarray}\normalsize
\noindent If $\boldsymbol{\hat{w}}=\boldsymbol{w}_{\mathrm{opt}}$, the SINR loss is maximum and is equal \mbox{to 1.} When we consider the LR structure of the noise, the theoretical SINR loss can be written as:\small
\begin{eqnarray}
  \rho_{\mathrm{LR}} &=&\frac{\vert\boldsymbol{w}_\mathrm{LR}^H\boldsymbol{d}\vert^2}{(\boldsymbol{w}_\mathrm{LR}^H\mathbf{R}\boldsymbol{w}_\mathrm{LR})(\boldsymbol{d}^H\mathbf{R}^{-1}\boldsymbol{d})}\\
  &=&\frac{\vert\boldsymbol{a}^H\boldsymbol{\Pi}_\mathrm{c}^{\bot}\boldsymbol{a}\vert^2}{(\boldsymbol{a}^H\boldsymbol{\Pi}_\mathrm{c}^{\bot}\mathbf{R}\boldsymbol{\Pi}_\mathrm{c}^{\bot}\boldsymbol{a})(\boldsymbol{a}^H\mathbf{R}^{-1}\boldsymbol{a})}
  \label{eq:SNRLoss_wLRopt}
\end{eqnarray}\normalsize
\noindent Finally, the SINR loss corresponding to the adaptive filter in Eq.(\ref{eq:wLRSCM}) is defined from Eq.(\ref{eq:SNRLoss_wLRopt}) as:
\begin{eqnarray}
  \hat{\rho}_{\mathrm{LR}}=\rho_{\mathrm{LR}}\vert_{\boldsymbol{\Pi}_\mathrm{c}^{\bot}= \hat{\boldsymbol{\Pi}}_{\mathrm{c}}^{\bot}}
  \label{eq:SNRLoss_wLRSCM}
\end{eqnarray}
\indent Since we are interested in the performance of the filters, we would like to obtain the theoretical behavior of the SINR losses. Some asymptotic studies on the SINR loss in LR Gaussian context have already been done~\cite{Ha97,GiFo13}. In~\cite{Ha97,GiFo13}, the theoretical result is derived by using the assumption that the LR noise is orthogonal to the steering vector and, in this case,~\cite{GiFo13} obtained an approximation of the expectation of the SINR loss $\hat{\rho}_\mathrm{LR}$. However, this assumption is not always verified, not very relevant and is a restrictive hypothesis in real cases. We consequently propose to relax it and study the convergence of the SINR loss using RMT tools through the study of the nominators and denominators. Indeed, one can already note that the numerators are \textit{simple} QFs whose convergences were widely considered in RMT. However, the denominators contain more elaborated QFs which were not tackled in RMT yet and will be the object of Sec.\ref{sec:NewCVResults}.

\section{Random matrix theory tools}\label{sec:RMT}
\indent This section is dedicated to the introduction of classical results from the RMT for the study of the convergence of QFs. This theory and the convergences are based on the behavior of the eigenvalues of the SCM when $m,K\rightarrow\infty$ at the same rate, i.e. $m/K\rightarrow c\in\left] 0,+\infty\right)$. In order to simplify the notations, we will abusively note $c=m/K$.\\
\indent The useful tools for the study of the eigenvalues behavior and the assumptions to the different convergences will be first presented. Secondly, the section will expose the data model, the \textit{spiked} model~\cite{CoHa13}. Finally, the useful convergences of \textit{simple} QFs ($\bm{s}_1^H\mathbf{\hat{R}}^{-1}\bm{s}_2$, $\bm{s}_1^H\bm{\hat{\Pi}}\bm{s}_2$) will be introduced.

\subsection{Preliminaries}
The asymptotic behavior of the eigenvalues when \mbox{$m,K\rightarrow\infty$} at the same rate is described through the convergence of their associated empirical Cumulative Distribution Function (CDF) $\hat{F}_m(x)$ or their empirical Probability Density Function (PDF) $\hat{f}_m(x)$\footnote{One can show that under (\textbf{As1},\textbf{As3}-\textbf{As5}) described later, $\hat{f}_m(x)$ a.s. converges towards a nonrandom PDF $f(x)$ with a compact support.}. The asymptotic PDF $f_m(x)$ will allow us to characterize the studied data model. The empirical CDF of the sample eigenvalues of $\mathbf{\hat{R}}$ can be \mbox{defined as:}
\begin{eqnarray}
	\hat{F}_m(x)=\frac{1}{m}\#\left\lbrace  k:\hat{\lambda}_k\leqslant x\right\rbrace
	\label{eq:Fhat}
\end{eqnarray}
\noindent However, in practice, the asymptotic characterization of $\hat{F}_m(x)$ is too hard. Consequently, one prefers to study the convergence of the Stieltjes transform ($\mathcal{ST}\left[ \cdot\right]$) of $\hat{F}_m(x)$:
\begin{eqnarray}
	\hat{b}_m(z)&=&\mathcal{ST}\left[ \hat{F}_m(x)\right]=\int_{\R}\dfrac{1}{x -z}d\hat{F}_m(x)\\
	&=&\dfrac{1}{m}\sum_{i=1}^m\dfrac{1}{\hat{\lambda}_i -z}=\dfrac{1}{m}\mathrm{tr}\left[(\mathbf{\hat{R}}-z\mathbf{I}_m)^{-1}\right] 
	\label{eq:ST_Fhat}
\end{eqnarray}
\noindent with $z\in\C^+\equiv\lbrace z\in\C :\Im[z]>0\rbrace$ and which almost surely converges to $\bar{b}_m(z)$. It is interesting to note that the PDF can thus be retrieve from the Stieltjes transform of its CDF:
\begin{eqnarray}
	\hat{f}_m(x)=\underset{\Im\left[z\right]\rightarrow 0 }{\mathrm{lim}}\frac{1}{\pi}\Im \left[\hat{b}_m(z)\right] 
	\label{eq:fhat}
\end{eqnarray}
\noindent with $x\in\R$. In an other manner, the characterization of $\hat{f}_m(x)$ (resp. $f_m(x)$) can be obtained from $\hat{b}_m(z)$ (resp. $\bar{b}_m(z)$). Then, to prove the convergences, we assume the following standard hypotheses.
\begin{itemize}[\labelsep =0.2cm]
\setlength{\itemindent}{0.3cm}
	\item[(\textbf{As1})] $\mathbf{R}$ has uniformly bounded spectral norm $\forall m\in\N^*$, i.e. \mbox{$\forall i\in[\![1,m]\!]$},  $\lambda_i < \infty$.
	\item[(\textbf{As2})] The vectors $\boldsymbol{s}_1$, $\boldsymbol{s}_2\in\C^{m\times 1}$ used in the QFs (here $\boldsymbol{a}(\bm{\Theta})$ and $\boldsymbol{x}$) have uniformly bounded Euclidean norm \mbox{$\forall m\in\N^*$}.
	\item[(\textbf{As3})] Let $\mathbf{Y}\in\C^{m\times K}$ having iid entries $y_{ij}$ with zero mean and unit variance, absolutely continuous and with $\E[|y_{ij}|^8]<\infty$.
\end{itemize}
\begin{itemize}[\labelsep =0.2cm]
\setlength{\itemindent}{0.3cm}
	\item[(\textbf{As4})] Let $\mathbf{Y}\in\C^{m\times K}$ defined as in (\textbf{As3}), then its distribution is invariant by left multiplication by a deterministic unitary matrix. Moreover, the eigenvalues empirical PDF of $\frac{1}{K}\mathbf{Y}\mathbf{Y}^H$ a.s. converges to the Mar\u{c}enko-Pastur distribution~\cite{MaPa67} with support $[(1-\sqrt{c})^2,(1+\sqrt{c})^2]$.
	\item[(\textbf{As5})] The maximum (resp. minimum) eigenvalue of $\frac{1}{K}\mathbf{Y}\mathbf{Y}^H$ a.s. tends to $(1+\sqrt{c})^2$ (resp. to $(1-\sqrt{c})^2$).
\end{itemize}
\subsection{Covariance matrix models and convergence of eigenvalues}\label{subsec:RMTmodels}
\begin{figure}[h!]
	\centering
	\includegraphics[scale=0.72]{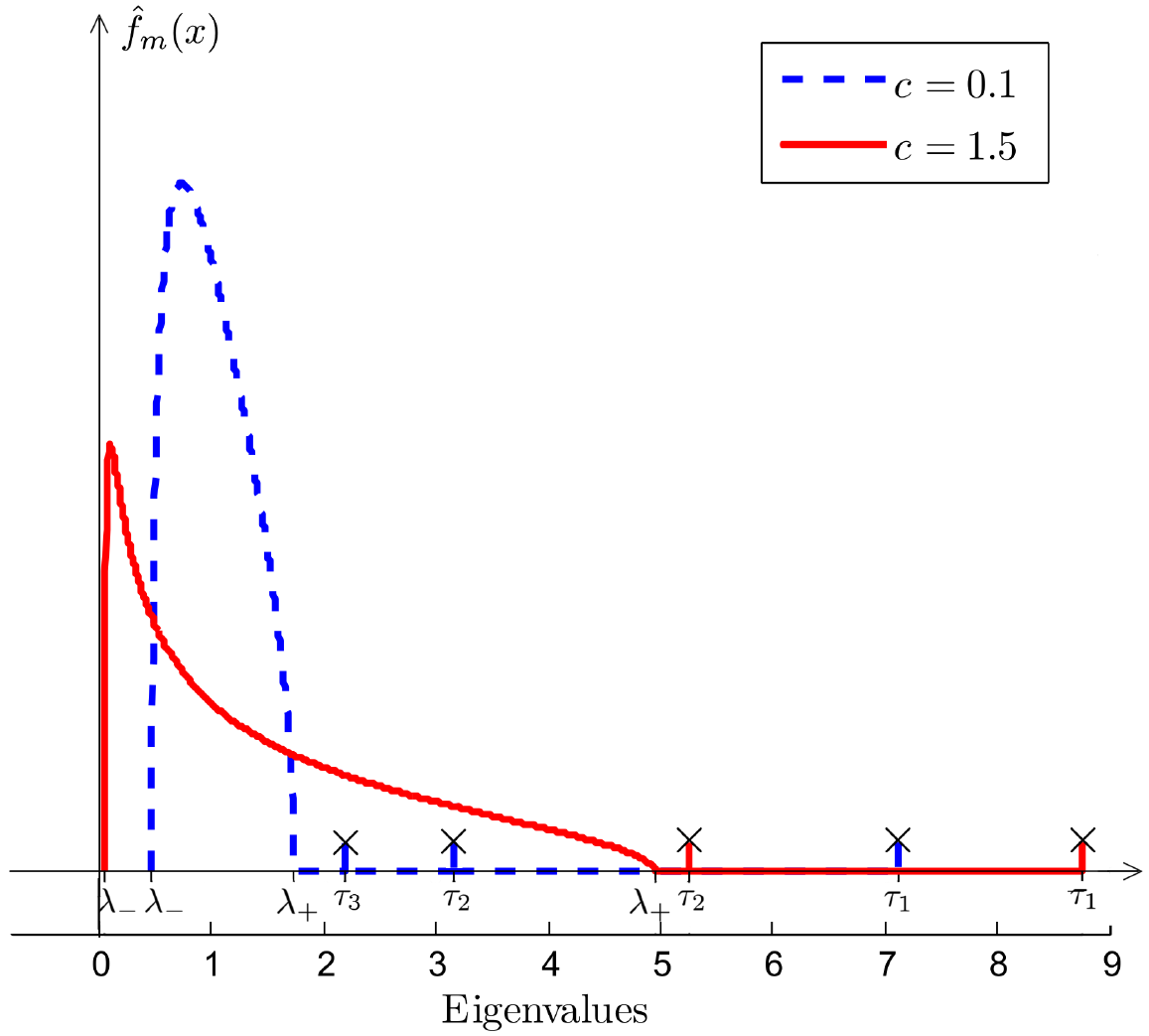}
	\caption{PDF of the eigenvalues of the SCM with the \textit{spiked} model when the eigenvalues of $\mathbf{R}$ are 1, 2, 3, and 7 with the same multiplicity, where 1 is the noise eigenvalue.}
	\label{Fig:ddpSPIKE_c01}
\end{figure}

\indent We first expose the considered data model and, then, the eigenvalues behavior of the SCM. The SCM can be written as $\mathbf{\hat{R}}=\frac{1}{K}\mathbf{X}\mathbf{X}^H$ with:
\begin{eqnarray}
  \mathbf{X}=\mathbf{R}^{1/2}\mathbf{Y}=(\mathbf{I}_m+\mathbf{P})^{1/2}\mathbf{Y}
  \label{eq:X_spike}
\end{eqnarray}
\noindent with $\mathbf{X}=[\boldsymbol{x}_1,\cdots,\boldsymbol{x}_K]$. The iid entries of $\mathbf{Y}$ follow the $\NC(0,1)$ distribution according to our data model in Sec.\ref{sec:pb_statement}. The complex normal distribution being a particular case of such distributions defined in (\textbf{As3}), the $\mathbf{Y}$ entries consequently verify it. Thus, the forthcoming convergences hold in the more general case defined by (\textbf{As3}). $\mathbf{R}^{1/2}$ is the $m\times m$ Hermitian positive definite square root of the true covariance matrix. The matrix $\mathbf{P}$ is the rank $r$ perturbation matrix and can be eigendecomposed as $\mathbf{P}=\mathbf{U}\bm{\Omega}\mathbf{U}^H =\sum_{i=1}^{\bar{M}}\omega_i\mathbf{U}_i\mathbf{U}_i^H$ with:
\begin{eqnarray}
  \bm{\Omega}=\begin{bmatrix}
  \omega_1 \mathbf{I}_{\mathcal{K}_1} & & \\
  &\ddots & \\
  & & \omega_{\bar{M}}\mathbf{I}_{\mathcal{K}_{\bar{M}}}
  \end{bmatrix}
  \label{eq:Omega}
\end{eqnarray}
\noindent with $\mathbf{U}=[\mathbf{U}_1\cdots\mathbf{U}_{\bar{M}}]$ and $\bar{M}$ the number of distinct eigenvalues of $\mathbf{R}$. Moreover, $\mathbf{U}_i\in\C^{m\times \mathcal{K}_i}$ where $\mathcal{K}_i$ is the multiplicity of $\omega_i$. Hence, the covariance matrix (\ref{eq:R}) can be rewritten as:
\begin{eqnarray}
  \mathbf{R}=\sum_{i=1}^{\bar{M}}\lambda_i\mathbf{U}_i\mathbf{U}_i^H
  \label{eq:Rgmusic}
\end{eqnarray}
\noindent where $\lambda_i$, of multiplicity $\mathcal{K}_i$, and $\mathbf{U}_i$ are the eigenvalues and the associated subspaces (concatenation of the $\mathcal{K}_i$ eigenvectors associated to $\lambda_i$) of $\mathbf{R}$ respectively, with $\lambda_1=1+\omega_1>\cdots>\lambda_ {\bar{M}}=1+\omega_{\bar{M}}>0$ and $\sum_{i=1}^{\bar{M}}\mathcal{K}_i=m$. The properties of the \textit{spiked} model are the following:
\begin{itemize}
	\item $\exists n\in[\![1,\bar{M}]\!]$ such that $\omega_n=0$.
	\item The multiplicity $\mathcal{K}_i$ is fixed $\forall i\in[\![1,\bar{M}]\!]\backslash n$ and does not increase with $m$, i.e. $\mathcal{K}_i/m\!\!\!\!\underset{m,K\rightarrow\infty}{\longrightarrow}\!\!\!\!0^+$, $\forall i\!\in\![\![1,\bar{M}]\!]\backslash n$.
\end{itemize}
\indent Consequently, we have $\mathrm{rank}(\mathbf{\Omega})=\sum_{i\in[\![1,\bar{M}]\!]\backslash n}\mathcal{K}_i=r$ and $\mathcal{K}_n=m-r$. In other words, the model specifies that only a few eigenvalues are non-unit (and do not contribute to the noise unit-eigenvalues) and fixed. Consequently, $\lambda_n=1$ is the eigenvalue of $\mathbf{R}$ corresponding to the white noise and the others correspond to the rank $r$ perturbation.\\ 
\indent In our case (see Sec.\ref{sec:pb_statement}), we recall that the covariance matrix $\mathbf{R}$ can be written as in Eq.(\ref{eq:R}) and Eq.(\ref{eq:Rgmusic}). More specifically, the noise component $\boldsymbol{b}$ corresponds to the white noise and its eigenvalue is $\lambda_{\bar{M}}=1$ as, for simplicity purposes, we set $\sigma^2=1$. The $r$ eigenvalues of the LR noise component $\boldsymbol{c}$ are strictly higher than 1. Thus, $\bar{M}=r+1$, $\lambda_1=1+\omega_1>\cdots>\lambda_{\bar{M}-1}=1+\omega_{\bar{M}-1}>\lambda_{\bar{M}}=1$, $\mathcal{K}_i=1$ is the multiplicity of $\lambda_i$, $\forall i\in[\![1,r]\!]$, and $\mathcal{K}_{\bar{M}}=m-r$ is the multiplicity of $\lambda_{\bar{M}}$.\small
\begin{eqnarray}
  \mathbf{R}\!=\!\lambda_{\bar{M}} \mathbf{U}_{\bar{M}}\mathbf{U}_{\bar{M}}^H\!+\!\!\sum_{i=1}^{\bar{M}-1}\lambda_i \mathbf{U}_i\mathbf{U}_i^H\!\!=\!\mathbf{U}_{r+1}\mathbf{U}_{r+1}^H\!+\!\!\sum_{i=1}^{r}\lambda_i \boldsymbol{u}_i\boldsymbol{u}_i^H\label{eq:SCM_spike}
\end{eqnarray} \normalsize
\indent This model leads to a specific asymptotic eigenvalues PDF of $\mathbf{R}$ as detailed hereafter. The convergence of the eigenvalues is addressed through the convergence of the Stieltjes transform of the eigenvalues CDF. The asymptotic eigenvalue behavior of $\hat{\mathbf{R}}$ for the \textit{spiked} model was introduced by Johnstone~\cite{Jo01} and its eigenvalue behavior was studied in~\cite{BaikSi06}. In order to derive it,~\cite{BaikSi06} exploited the specific expression given in Eq.(\ref{eq:X_spike}). Then,~\cite{CoHa13} introduced the final assumption (\textit{separation condition}) under which the following convergences are given.
\begin{itemize}[\labelsep =0.2cm]
\setlength{\itemindent}{0.55cm}
	\item[(\textbf{As6.S})] The eigenvalues of $\mathbf{P}$ satisfy the \textit{separation condition}, i.e. $\vert\omega_i\vert >\sqrt{c}$ for all $i\in[\![1,\bar{M}]\!]\backslash n$ ($i\in[\![1,r]\!]$ in our case).
\end{itemize}
\noindent Thus, under (\textbf{As1}-\textbf{As5}, \textbf{As6.S}), we have:
\begin{eqnarray}
	\hat{f}_m(x)\longrightarrow f(x)
\end{eqnarray}
\noindent where $f(x)$ is the Mar\u{c}enko-Pastur law:
\begin{eqnarray}
	f(x)=\begin{cases}\left(1-\frac{1}{c}\right),\qquad\qquad\text{ if } x=0\text{ and }c>1\\
	\dfrac{1}{2\pi cx}\sqrt{(\lambda_- -x)(x -\lambda_+)},\\
	\qquad\qquad\qquad\qquad\:\:\!\text{if } x \in]\lambda_-,\lambda_+[\\
	0, \qquad\qquad\qquad\quad\:\:\text{otherwise}
	\end{cases}
\end{eqnarray}
\noindent with $\lambda_-=(1-\sqrt{c})^2$ and $\lambda_+=(1+\sqrt{c})^2$. However, it is essential to note that, for all $i\in[\![1,\bar{M}]\!]\backslash n$: 
\begin{eqnarray}
	\hat{\lambda}_{j\in\mathcal{M}_i}\overset{\mathrm{a.s.}}{\underset{m,K\rightarrow\infty}{\longrightarrow}}\tau_i=1+\omega_i+c\frac{1+\omega_i}{\omega_i}\label{eq:rho}
\end{eqnarray} 
\noindent where $\mathcal{M}_i$ is the set indexes corresponding to the $j$-th eigenvalue of $\mathbf{R}$ (for example $\mathcal{M}_{r+1}=\left\lbrace r+1,\cdots,m\right\rbrace $ for $\lambda_{r+1}$). Two representations of $\hat{f}_m(x)$ for two different $c$ and a sufficient large $m$ are shown on Fig.~\ref{Fig:ddpSPIKE_c01} when the eigenvalues of $\mathbf{R}$ are 1, 2, 3, and 7 with the same multiplicity, where the eigenvalue 1 is the noise eigenvalue. One can observe that say (\textbf{As6.S}) is verified is equivalent to say that $\tau_{n-1}>\lambda_+$ and $\tau_{n+1}<\lambda_-$. In other words, all the sample eigenvalues corresponding to the non-unit eigenvalues of $\mathbf{R}$, converge to a value $\tau_i$ which is outside the support of the Mar\u{c}enko-Pastur law (\enquote{asymptotic} PDF of the \enquote{unit} sample eigenvalues). As an illustration, one can notice that, in Fig.~\ref{Fig:ddpSPIKE_c01}, for $\hat f_m(x)$ plotted for $c=0.1$, the \textit{separation condition} is verified ($\omega_1=6$, $\omega_2=2$ and $\omega_3=1$ are greater that $\sqrt{c}=0.316$) and the three non-unit eigenvalues are represented on the PDF and outside the support of the Mar\u{c}enko-Pastur law by their respective limits $\tau_1=7.116$, $\tau_2=3.15$ and $\tau_3=2.2$. On the contrary, for $\hat{f}_m(x)$ plotted for $c=1.5$, only the two greatest eigenvalues are represented on the PDF by their respective limits $\tau_1=8.75$ and $\tau_2=5.25$ while the \textit{separation condition} is not verified for the eigenvalue $\lambda_3=2$ ($\omega_3=1<\sqrt{c}=1.223$). In this case, the sample eigenvalues corresponding to the eigenvalue $\lambda_3=2$ belongs to the Mar\u{c}enko-Pastur law.
\subsection{Convergence of simple quadratic forms}
\indent Here, we compare the convergence of two QFs in two convergence regimes: when \mbox{$K\rightarrow\infty$} with a fixed $m$ and when $m,K\rightarrow\infty$ at the same rate. \\
\indent We first present the useful convergences of \textit{simple} QFs function of $\mathbf{\hat{R}}$. It is well known that, due to the strong law of large numbers, when \mbox{$K\rightarrow\infty$} with a fixed $m$, $\mathbf{\hat{R}}\rightarrow\mathbf{R}$ a.s.~\cite{Bi95}. Thus,
\begin{eqnarray}
	\boldsymbol{s}_1^H\mathbf{\hat{R}}^{-1}\boldsymbol{s}_2 \underset{\underset{m<\infty}{\small{K\rightarrow\infty}}}{\overset{\text{a.s.}}{\longrightarrow}} \boldsymbol{s}_1^H \mathbf{R}^{-1}\boldsymbol{s}_2
	\label{eq:CV_FQR}
\end{eqnarray}
\noindent Moreover, when $m,K\rightarrow\infty$ at the same rate~\cite{Gi98,Me08}:
\begin{eqnarray}
	\boldsymbol{s}_1^H\mathbf{\hat{R}}^{-1}\boldsymbol{s}_2 \underset{\underset{m/K \to c<\infty}{\small{m,K\rightarrow\infty}}}{\overset{\text{a.s.}}{\longrightarrow}} \left(1-c\right)^{-1}\boldsymbol{s}_1^H \mathbf{R}^{-1}\boldsymbol{s}_2
	\label{eq:CV_FQR_RMT}
\end{eqnarray}
\indent The useful convergences of \textit{simple} QFs function of $\hat{\boldsymbol{\Pi}}_{\mathrm{c}}^{\bot}$ are then presented. As $\mathbf{\hat{R}}\rightarrow\mathbf{R}$ a.s. when \mbox{$K\rightarrow\infty$} with a fixed $m$, $\hat{\boldsymbol{\Pi}}_{\mathrm{c}}^{\bot}\rightarrow\boldsymbol{\Pi}_{\mathrm{c}}^{\bot}$ a.s.~\cite{Me08} in the same convergence regime. Thus:
\begin{eqnarray}
	\boldsymbol{s}_1^H\hat{\boldsymbol{\Pi}}_{\mathrm{c}}^{\bot}\boldsymbol{s}_2 \:\underset{\underset{m<\infty}{\small{K\rightarrow\infty}}}{\overset{\text{a.s.}}{\longrightarrow}} \,\boldsymbol{s}_1^H\boldsymbol{\Pi}_{\mathrm{c}}^{\bot}\boldsymbol{s}_2\label{eq:CV_FQsimple_Kinf}
\end{eqnarray}
\indent For the convergences in the large dimensional regime ($m,K\rightarrow\infty$ at the same rate), the convergences are presented under (\textbf{As1}-\textbf{As5}) and the \textit{separation condition} \textbf{As6.S}. \cite{CoHa13} showed that, $\forall i\in[\![1,\bar{M}-1]\!]$:
\begin{eqnarray}
    \boldsymbol{s}_1^H\mathbf{\hat{U}}_i\mathbf{\hat{U}}_i^H\boldsymbol{s}_2 \underset{\underset{m/K \to c<\infty}{\small{m,K\rightarrow\infty}}}{\overset{\text{a.s.}}{\longrightarrow}}  \dfrac{1-c\omega_i^{-2}}{1+c\omega_i^{-1}}\boldsymbol{s}_1^H\mathbf{U}_i\mathbf{U}_i^H\boldsymbol{s}_2
\end{eqnarray}
\noindent with $\omega_i=\lambda_i-1$. $\lambda_i$ is the $i$-th distinct eigenvalue of $\mathbf{R}$. Let $\chi_i=\dfrac{1-c\omega_i^{-2}}{1+c\omega_i^{-1}}$. Thus, using the following relationship,
\begin{eqnarray}
    \hat{\boldsymbol{\Pi}}_{\mathrm{c}}^{\bot}=\mathbf{I}_m-\sum_{i=1}^{\bar{M}-1}\mathbf{\hat{U}}_i\mathbf{\hat{U}}_i^H=\mathbf{I}_m-\sum_{i=1}^{r}\boldsymbol{\hat{u}}_i\boldsymbol{\hat{u}}_i^H
\end{eqnarray}
one can deduce that with the \textit{spiked} model and in the large dimensional regime:
\begin{eqnarray}
	\boldsymbol{s}_1^H\hat{\boldsymbol{\Pi}}_{\mathrm{c}}^{\bot}\boldsymbol{s}_2 \underset{\underset{m/K \to c<\infty}{\small{m,K\rightarrow\infty}}}{\overset{\text{a.s.}}{\longrightarrow}}  \boldsymbol{s}_1^H\bar{\boldsymbol{\Pi}}_{\mathrm{c,S}}^{\bot}\boldsymbol{s}_2
    \label{eq:CV_LRFQ_SpikeSCM}
\end{eqnarray}
\noindent with $\bar{\boldsymbol{\Pi}}_{\mathrm{c,S}}^{\bot}=\sum_{i=1}^{m} \psi_i\boldsymbol{u}_i\boldsymbol{u}_i^H$ and
\begin{eqnarray}
	\psi_i=\begin{cases}1,\;\,\qquad\quad \mathrm{if}\;i>r\\
		1-\chi_i,\;\quad \mathrm{if}\;i\leqslant r
	\end{cases}\label{eq:PiOrthS}
\end{eqnarray}
\indent Consequently, $\boldsymbol{s}_1^H\hat{\mathbf{R}}^{-1}\boldsymbol{s}_2$ is consistent in the two convergence regimes and, although $\boldsymbol{s}_1^H\hat{\boldsymbol{\Pi}}_{\mathrm{c}}^{\bot}\boldsymbol{s}_2$ is consistent when $K\rightarrow\infty$ with a fixed $m$, it is no more consistent under the regime of interest i.e. when both $m,K\rightarrow\infty$ at the same rate.

\section{New convergence results}\label{sec:NewCVResults}
\subsection{Convergence of structured quadratic forms}
\indent In this section, the convergence of the \textit{structured} QF function of $\hat{\boldsymbol{\Pi}}_{\mathrm{c}}^{\bot}$ is analyzed and results to Proposition 1.\\
\indent\textit{\textbf{Proposition 1:}} Let $\mathbf{B}$ be a $m\times m$ deterministic complex matrix with a uniformly bounded spectral norm for all $m$. Then, under (\textbf{As1}-\textbf{As5}, \textbf{As6.S}) and the \textit{spiked} model,
\begin{eqnarray}
	\begin{array}{l}
		\boldsymbol{s}_1^H\hat{\boldsymbol{\Pi}}_{\mathrm{c}}^{\bot}\mathbf{B}\hat{\boldsymbol{\Pi}}_{\mathrm{c}}^{\bot}\boldsymbol{s}_2\underset{\underset{m/K \to c<\infty}{\small{m,K\rightarrow\infty}}}{\overset{\text{a.s.}}{\longrightarrow}}  \boldsymbol{s}_1^H\bar{\boldsymbol{\Pi}}_{\mathrm{c,S}}^{\bot}\mathbf{B}\bar{\boldsymbol{\Pi}}_{\mathrm{c,S}}^{\bot}\boldsymbol{s}_2
	\end{array}\label{eq:CV_LRFQ2_SpikeSCM}
\end{eqnarray}
where $\bar{\boldsymbol{\Pi}}_{\mathrm{c,S}}^{\bot}=\sum_{i=1}^{m} \psi_i\boldsymbol{u}_i\boldsymbol{u}_i^H$ with $\psi_i$ defined by Eq.(\ref{eq:PiOrthS}).
\begin{flushright} \vspace{-0.3cm}$\blacksquare$ \end{flushright}

\begin{figure}[h!]
	\centering
	\includegraphics[scale=0.6]{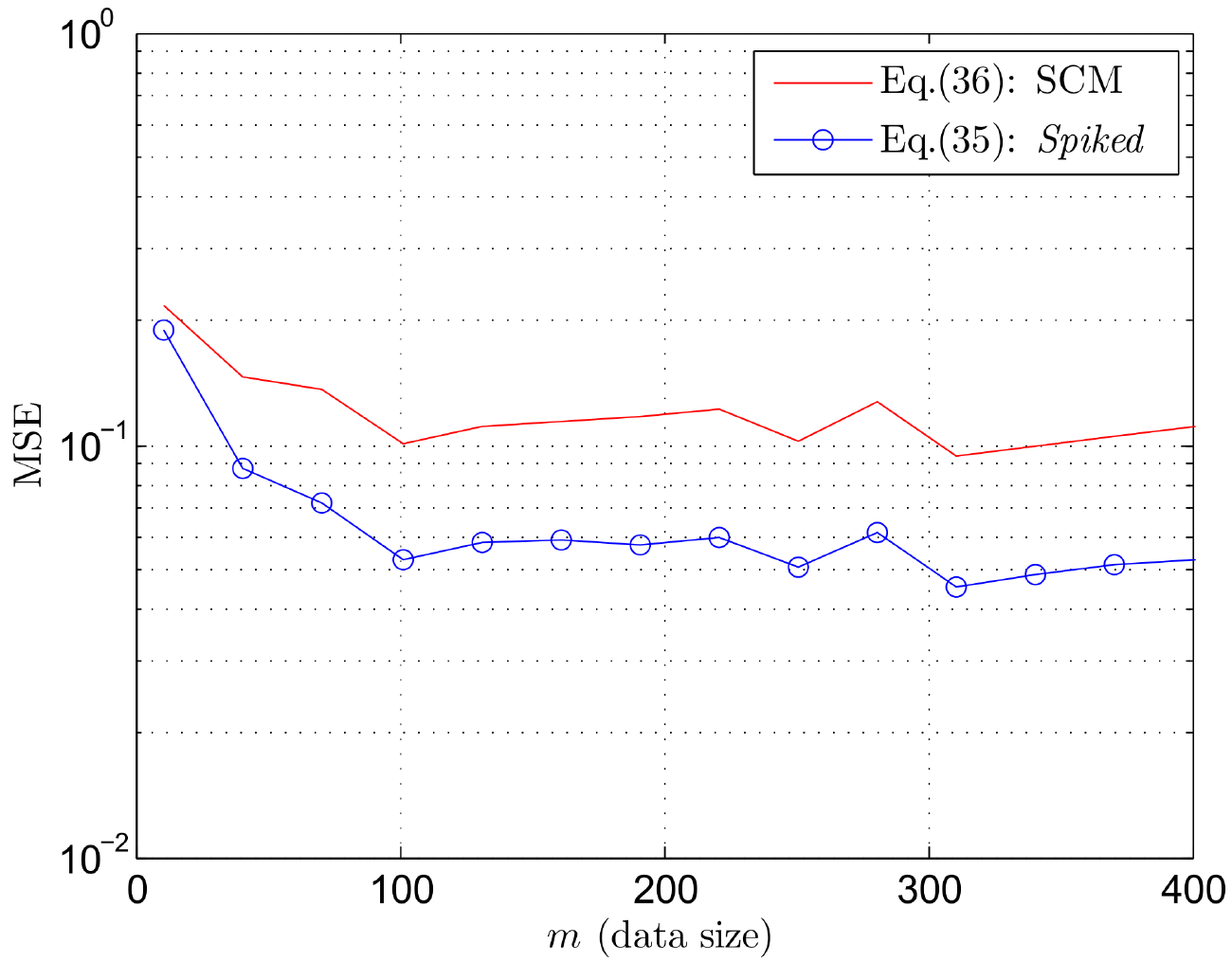}
	\caption{MSE over $10^3$ iterations corresponding to Eq.(\ref{eq:CV_LRFQ2_SpikeSCM}) and Eq.(\ref{eq:CV_LRFQ2_SCM}) when the eigenvalues of $\mathbf{R}$ are 1, 21, 31, and 71 with the multiplicity $m-3$, 1, 1 and 1 respectively, $c=0.1$, $\boldsymbol{s}_1=\boldsymbol{s}_2$ are steering vectors of the LR noise component $\boldsymbol{c}$ and $\mathbf{B}=\mathbf{R}$.}
	\label{Fig:lemma3}
\end{figure}

\indent \textit{Proof:} See Appendix.\\\\
\noindent Moreover, one can remark that if $\mathbf{B}=\mathbf{R}$, where $\mathbf{R}$ is the covariance matrix as defined in Eq.(\ref{eq:Rscm}), the following convergence holds:
\begin{eqnarray}
	\begin{array}{l}
		\boldsymbol{s}_1^H\hat{\boldsymbol{\Pi}}_{\mathrm{c}}^{\bot}\mathbf{R}\hat{\boldsymbol{\Pi}}_{\mathrm{c}}^{\bot}\boldsymbol{s}_2\underset{\underset{m/K \to c<\infty}{\small{m,K\rightarrow\infty}}}{\overset{\text{a.s.}}{\longrightarrow}}  \boldsymbol{s}_1^H\bar{\boldsymbol{\Pi}}_{\mathrm{c,S}}^{\bot}\mathbf{R}\bar{\boldsymbol{\Pi}}_{\mathrm{c,S}}^{\bot}\boldsymbol{s}_2
	\end{array}
\end{eqnarray}
\noindent A visualization of the convergence of Eq.(\ref{eq:CV_LRFQ2_SpikeSCM}) in terms of Mean Squared Error (MSE) can be found in Fig.~\ref{Fig:lemma3} when $m,K\rightarrow\infty$ at a fixed ratio. It is compared to the MSE corresponding to the following convergence when $K\rightarrow\infty$ with a fixed $m$:
\begin{eqnarray}
	\begin{array}{l}
		\boldsymbol{s}_1^H\hat{\boldsymbol{\Pi}}_{\mathrm{c}}^{\bot}\mathbf{B}\hat{\boldsymbol{\Pi}}_{\mathrm{c}}^{\bot}\boldsymbol{s}_2\underset{\underset{m<\infty}{\small{K\rightarrow\infty}}}{\overset{\text{a.s.}}{\longrightarrow}}  \boldsymbol{s}_1^H\boldsymbol{\Pi}_{\mathrm{c}}^{\bot}\mathbf{B}\boldsymbol{\Pi}_{\mathrm{c}}^{\bot}\boldsymbol{s}_2
	\end{array}\label{eq:CV_LRFQ2_SCM}
\end{eqnarray}

\subsection{Convergence of SINR losses}
\indent Now, we provide the convergences of the estimated SINR loss using the convergences previously presented and the following convergence. We recall that, as $\hat{\mathbf{R}}\rightarrow\mathbf{R}$ a.s. when $K\rightarrow\infty$ with a fixed $m$, one has:
\begin{eqnarray}
	\boldsymbol{s}_1^H\mathbf{\hat{R}}^{-1}\mathbf{R}\mathbf{\hat{R}}^{-1}\boldsymbol{s}_2 \underset{\underset{m<\infty}{\small{K\rightarrow\infty}}}{\overset{\text{a.s.}}{\longrightarrow}} \boldsymbol{s}_1^H \mathbf{R}^{-1}\boldsymbol{s}_2\label{eq:CV_FQR2}
\end{eqnarray}
Hence, when $K\rightarrow\infty$ with a fixed $m$ and using Eq.(\ref{eq:CV_FQR}), Eq.(\ref{eq:CV_FQR2}) and the continuous mapping theorem~\cite{Bi95}: 
\begin{eqnarray}
  \hat{\rho}\underset{\underset{m<\infty}{\small{K\rightarrow\infty}}}{\overset{\text{a.s.}}{\longrightarrow}}\frac{\vert\boldsymbol{a}^H\mathbf{R}^{-1}\boldsymbol{a}\vert^2}{(\boldsymbol{a}^H\mathbf{R}^{-1}\boldsymbol{a})(\boldsymbol{a}^H\mathbf{R}^{-1}\boldsymbol{a})}=1
\end{eqnarray}
And, under (\textbf{As1}-\textbf{As5}), when $m,K\rightarrow\infty$ at the same rate, from~\cite{TaTaPe10}, we have:
\begin{eqnarray}\small
  \hat{\rho}\underset{\underset{m/K \to c<\infty}{\small{m,K\rightarrow\infty}}}{\overset{\text{a.s.}}{\longrightarrow}}\!\!\frac{(1-c)\vert\boldsymbol{a}^H\mathbf{R}^{-1}\boldsymbol{a}\vert^2}{(\boldsymbol{a}^H\mathbf{R}^{-1}\boldsymbol{a})(\boldsymbol{a}^H\mathbf{R}^{-1}\boldsymbol{a})}=1-c
  \label{eq:SNRLossopt}
\end{eqnarray}\normalsize
\noindent Thus, the estimated SINR loss $\hat{\rho}$ is consistent when $K\rightarrow\infty$ with $m$ fixed and when $m,K\rightarrow\infty$ at the same rate, up to an additive constant $c$. Consequently, RMT cannot help us to improve the estimation of the theoretical SINR loss.\\
\indent For the SINR loss corresponding to the adaptive LR filters, when $K\rightarrow\infty$ with a fixed $m$, using Eq.(\ref{eq:CV_FQsimple_Kinf}), Eq.(\ref{eq:CV_LRFQ2_SCM}) and the continuous mapping theorem theorem, we have:
\begin{eqnarray}
	\hat{\rho}_{\mathrm{LR}}\underset{\underset{m<\infty}{\small{K\rightarrow\infty}}}{\overset{\text{a.s.}}{\longrightarrow}}\rho_{\mathrm{LR}}
  \label{eq:CV_SNRLoss_Kinf}
\end{eqnarray}
\noindent where $\rho_{\mathrm{LR}}$ is defined by Eq.(\ref{eq:SNRLoss_wLRopt}). When $m,K\rightarrow\infty$ at the same rate, we obtain the following convergence:
\begin{eqnarray}
		\hat{\rho}_{\mathrm{LR}}\underset{\underset{m/K \to c<\infty}{\small{m,K\rightarrow\infty}}}{\overset{\text{a.s.}}{\longrightarrow}}\bar{\rho}_{\mathrm{LR}}^{(\mathrm{S})}=\rho_{\mathrm{LR}}\vert_{\boldsymbol{\Pi}_\mathrm{c}^{\bot}= \boldsymbol{\bar{\Pi}}_{\mathrm{c},\mathrm{S}}^{\bot}}\neq\rho_{\mathrm{LR}}
  \label{eq:CV_SNRLoss_Spike}
\end{eqnarray}
\noindent where Eq.(\ref{eq:CV_LRFQ_SpikeSCM}), Proposition 1 and the continuous mapping theorem were used to prove Eq.(\ref{eq:CV_SNRLoss_Spike}). One can observe that, although the traditional estimator of $\rho_{\mathrm{LR}}$ is consistent when $K\rightarrow\infty$ with a fixed $m$, it is no more consistent when $m,K\rightarrow\infty$ at the same rate. It is also important to underline that the new convergence result leads to a more precise approximation of $\hat{\rho}_{\mathrm{LR}}$ than previous works~\cite{GiFo13}. Indeed,~\cite{GiFo13} proposes an approximation dependent on $K$ and the approximation proposed here depends on $K$ (and of course on $c$) as well as on the parameter $\bm{\Theta}$. 

\section{Simulations}\label{sec:simu}
\subsection{Parameters}
\begin{figure}[h!]
	\centering
	\includegraphics[scale=0.6]{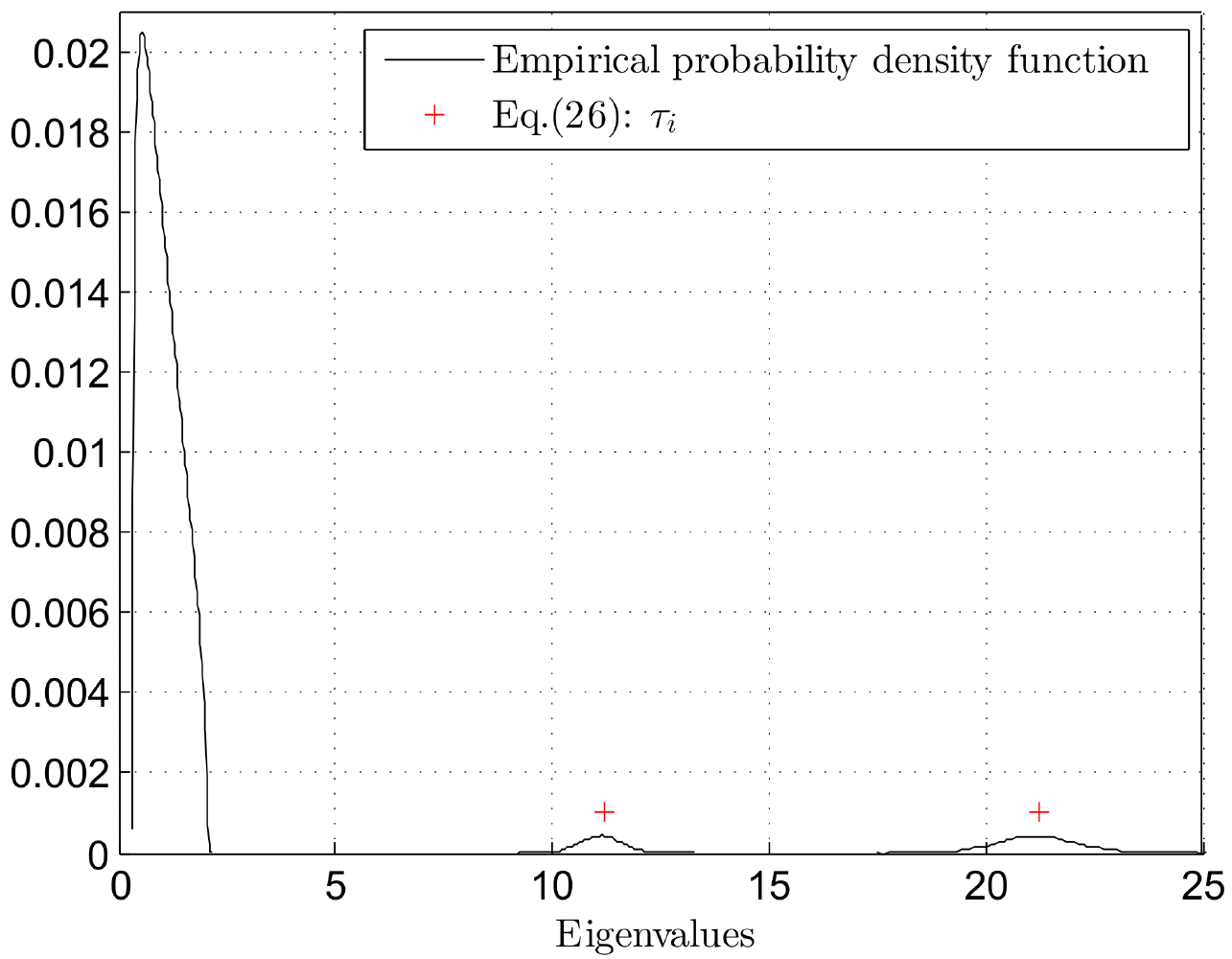}
	\caption{Zoom of the experimental PDF of jamming plus noise data with $c=0.2$ and $\frac{JNR}{\mathrm{tr}(\bm{\Lambda})}=10$dB.}
	\label{Fig:ddp_STAP}
\end{figure}

\begin{figure}[h!]
	\centering
	\includegraphics[scale=0.6]{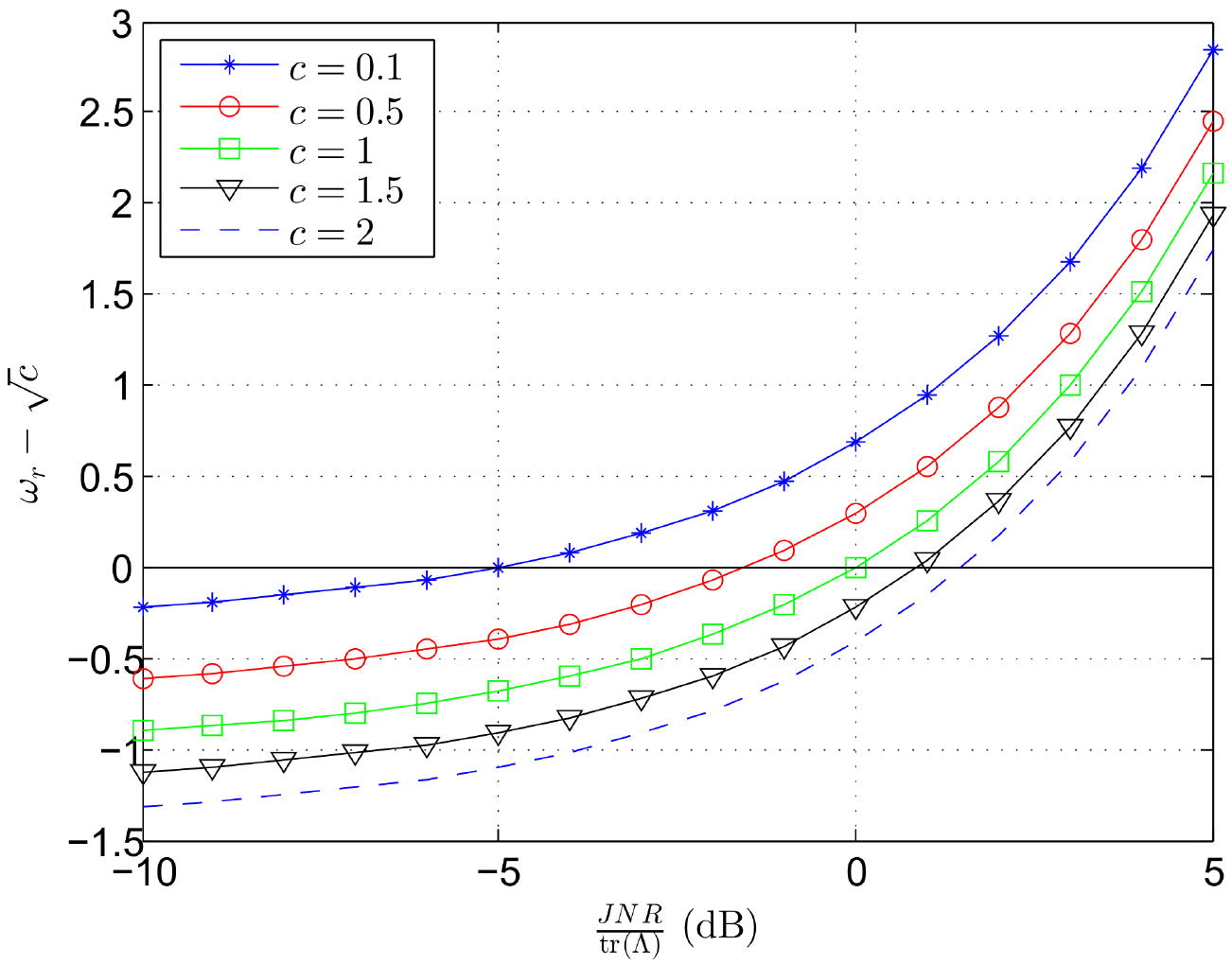}
	\caption{\textit{Separation condition} ($\omega_r-\sqrt{c}$) of the \textit{spiked} model for the lowest non-unit eigenvalue as a function of the ratio $\frac{JNR}{\mathrm{tr}(\bm{\Lambda})}$ in dB.}
	\label{Fig:separationSPIKE_rBrennan}
\end{figure}

\indent As an illustration of the interest of the application of RMT in filtering, the jamming application is chosen. The purpose of this application is to detect a target thanks to a ULA composed of $m$ sensors despite the presence of jamming. The response of the jamming, $\boldsymbol{c}$ is composed of signals similar to the target response. In this section, except for the convergences when $m,K\rightarrow\infty$ at the same rate $c$, we choose $m=100$ in order to have a large number for the data dimension. Even if, in some basic array processing applications, this number could seem significant, it actually became standard in many applications such as STAP~\cite{Wa94}, MIMO applications~\cite{LiSt09,TsVi05}, MIMO-STAP~\cite{LiSt09}, etc. Here, $\bm{\Theta}=\theta$ where $\theta$ is the AoA. The jamming is composed of three synthetic targets with AoA $-20^\circ$, $0^\circ$ and $20^\circ$ and wavelength $l_0=0.667$m. Thus, the jamming (LR noise) has a rank $r=3$. Then, the AWGN $\boldsymbol{b}$ power is $\sigma^2=1$. Finally, the theoretical covariance matrix of the total noise can be written as $\mathbf{R}=\frac{JNR}{\mathrm{tr}(\bm{\Lambda})}\mathbf{U}\bm{\Lambda}\mathbf{U}^H+\sigma^2\mathbf{I}_m$ with $\bm{\Lambda}=\mathrm{diag([6,2,1])}$ and where $JNR$ is the jamming to noise ratio. $\frac{JNR}{\mathrm{tr}(\bm{\Lambda})}$ is fixed at $10$dB except for Fig. 4.\\
\indent In order to validate the \textit{spiked} model as covariance matrix model, we visualize a zoom of the experimental PDF of the eigenvalues of our data without target in Fig.~\ref{Fig:ddp_STAP} over $5\times 10^4$ Monte-Carlo iterations. We observe a Mar\u{c}enko-Pastur law around 1 (eigenvalues of the white noise) and Gaussian distributions for the eigenvalues of the jamming, which is consistent to the CLT for the \textit{spiked} model proved in~\cite{CoHa13}. The \textit{spiked} model is consequently relevant for our data model.\\
\indent Moreover, in order to verified that the \textit{spiked} model is realistic in terms of \textit{separation condition}, Fig.~\ref{Fig:separationSPIKE_rBrennan} shows ($\omega_r-\sqrt{c}$) as a function of $\frac{JNR}{\mathrm{tr}(\bm{\Lambda})}$ in dB. This figure will be the same for all $m$ and $K$ at a fixed ratio. We recall that, in order to satisfy the \textit{separation condition}, one should have $\omega_r-\sqrt{c}>0$. Consequently, we gladly observe that it is satisfied for $\frac{JNR}{\mathrm{tr}(\bm{\Lambda})}>4$dB for the majority of $c$ even $c>2$. Indeed, in practice, if the $\frac{JNR}{\mathrm{tr}(\bm{\Lambda})}$ is lower, the jamming will not have any effects on the performance. 

\subsection{Performance of filters}
\begin{figure}[h!]
	\centering
	\includegraphics[scale=0.6]{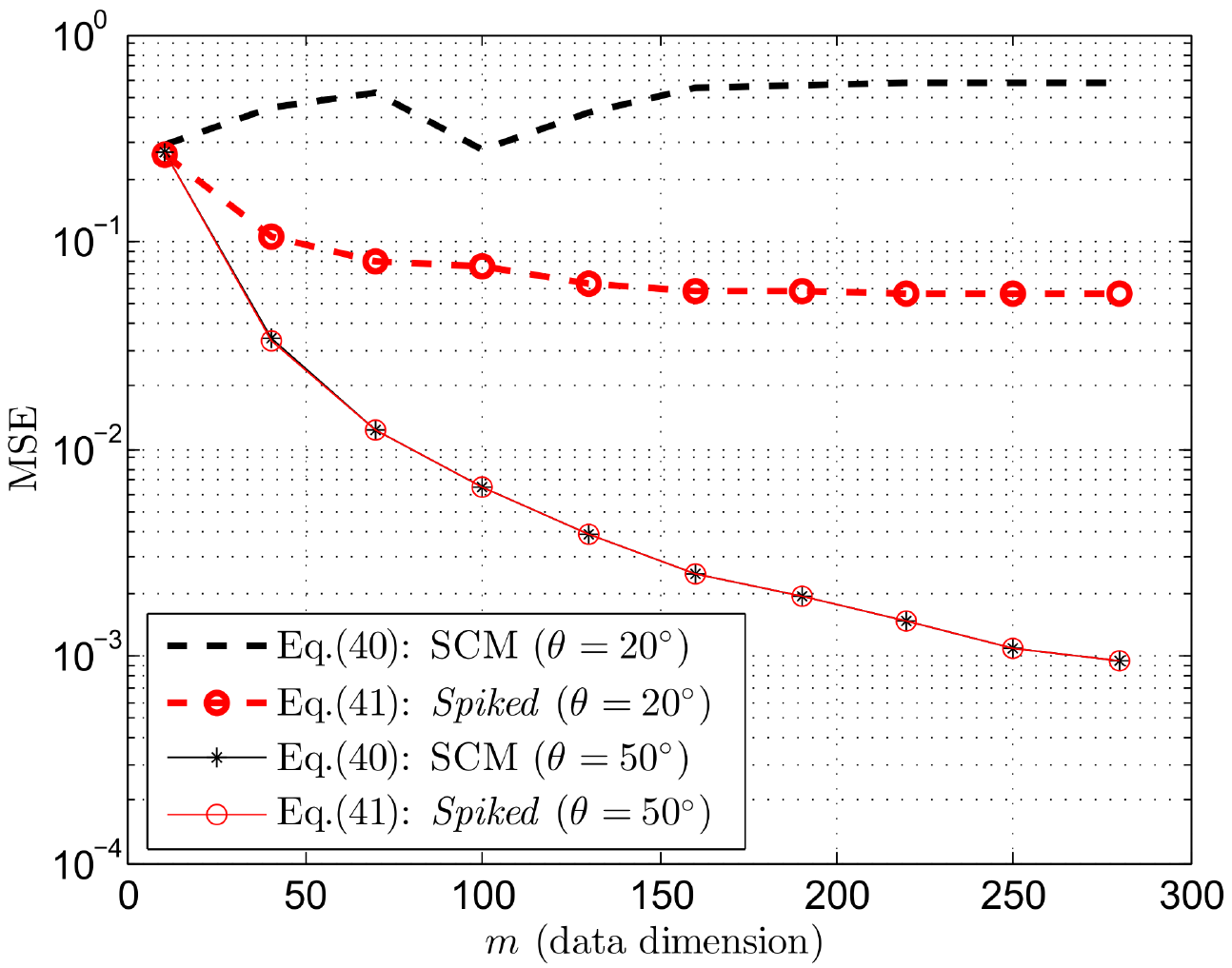}
	\caption{MSE corresponding to Eq.(\ref{eq:CV_SNRLoss_Kinf}) and Eq.(\ref{eq:CV_SNRLoss_Spike}) when $m,K\rightarrow\infty$ at a fixed ratio $c=3$ and $\frac{JNR}{\mathrm{tr}(\bm{\Lambda})}=10$dB.}
	\label{Fig:CV_SNRLoss_mKinf}
\end{figure}

\begin{figure}[h!]
	\centering
	\includegraphics[scale=0.63]{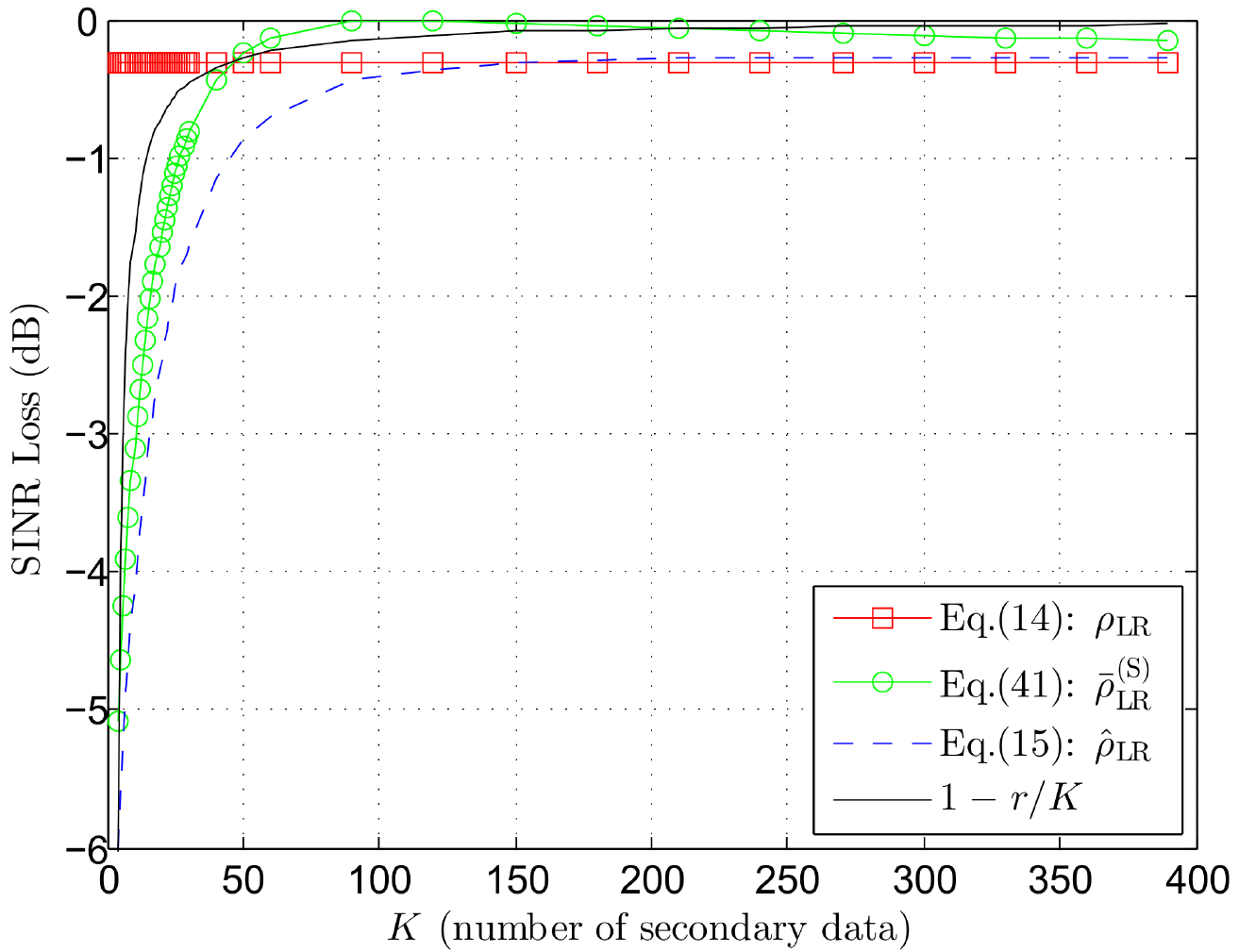}
	\caption{Visualization of Eq.(\ref{eq:SNRLoss_wLRopt}) (red line with squares), Eq.(\ref{eq:SNRLoss_wLRSCM}) (blue dashed line), the right side of the convergence in Eq.(\ref{eq:CV_SNRLoss_Spike}) (green line with circles) and the traditional estimation of $\E[\hat{\rho}_\mathrm{LR}]$ (black line) as a function of $K$ (over $10^3$ realizations) with $\frac{JNR}{\mathrm{tr}(\bm{\Lambda})}=10$dB, $m=100$ and $\theta=20.5^\circ$.}
	\label{Fig:SINRLoss_Kinf_theta205_avec}
\end{figure}

\begin{figure}[h!]
	\centering
	\includegraphics[scale=0.63]{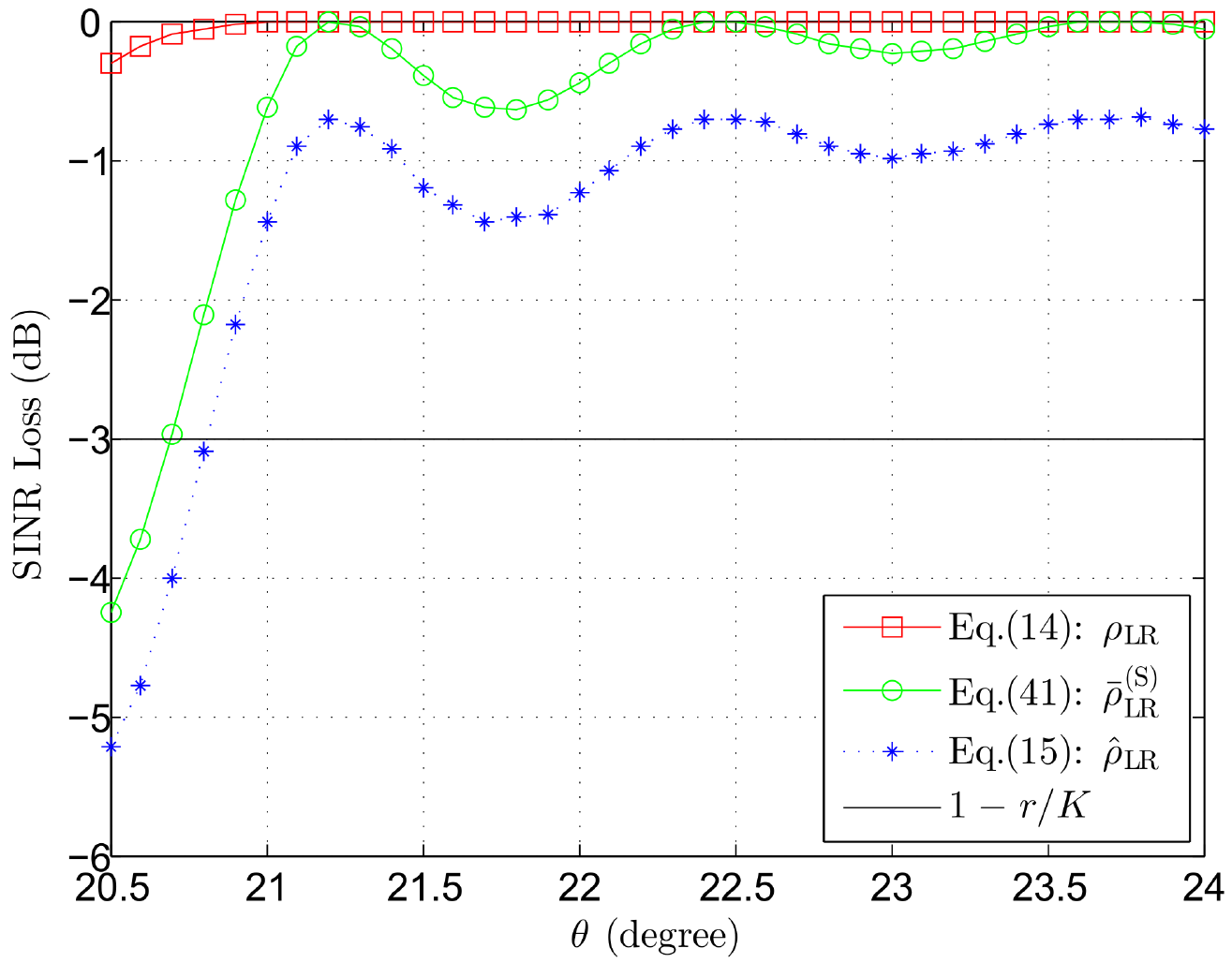}
	\caption{Visualization of Eq.(\ref{eq:SNRLoss_wLRopt}) (red line with squares), Eq.(\ref{eq:SNRLoss_wLRSCM}) (blue dashed line), the right side of the convergence in Eq.(\ref{eq:CV_SNRLoss_Spike}) (green line with circles) and the traditional estimation of $\E[\hat{\rho}_\mathrm{LR}]$ (black line) as a function of $\theta$ (over $10^3$ realizations) with $\frac{JNR}{\mathrm{tr}(\bm{\Lambda})}=10$dB, $m=100$ and $K=2r$.}
	\label{Fig:SNRloss_theta_CNR10_K2r}
\end{figure}

\indent We now observe the performances of filters through the SINR loss. We are first interested in the validation of the convergence of $\hat{\rho}_{\mathrm{LR}}$ in Eq.(\ref{eq:CV_SNRLoss_Spike}) as $m,K\rightarrow\infty$ at the same rate. This convergence is validated and presented in Fig.~\ref{Fig:CV_SNRLoss_mKinf} in terms of MSE over $10^3$ realizations with $c=3$ for an AoA of the target ($\theta=50^\circ$) and an AoA of the jamming ($\theta=20^\circ$).
\\
\indent Fig.~\ref{Fig:SINRLoss_Kinf_theta205_avec} shows the visualization of Eq.(\ref{eq:SNRLoss_wLRopt}) (blue line with stars), Eq.(\ref{eq:SNRLoss_wLRSCM}) (blue dashed line), the right side of the convergence in Eq.(\ref{eq:CV_SNRLoss_Spike}) (green line with circles) and the approximation $\E[\hat{\rho}_\mathrm{LR}]\simeq 1-\frac{r}{K}$ introduced by~\cite{GiFo13} (black line) as a function of $K$ when the target is near from the jamming, i.e. $\theta=20.5^\circ$. We observe that the \textit{spiked} model and the RMT helps us to obtain a better estimation of $\E[\hat{\rho}_\mathrm{LR}]$ than the estimation $\E[\hat{\rho}_\mathrm{LR}]\simeq 1-\frac{r}{K}$ as the curve of $\bar{\rho}_\mathrm{LR}^{(\mathrm{S})}$ has the same behavior as the curve of $\hat{\rho}_\mathrm{LR}$. Then, similarly, the same equations are visualized as a function of $\theta$ in Fig.~\ref{Fig:SNRloss_theta_CNR10_K2r} with $K=2r$. We observe that, unlike the estimation $1-r/K$, the RMT with the \textit{spiked} model permits us to obtain a better estimation of $\E[\hat{\rho}_\mathrm{LR}]$ as a function of $\theta$ and consequently a better approximation of its behavior. Thus, it permits to predict the parameter $\theta$ value corresponding to the performance break (here around $21.1^\circ$).

\section{Conclusion}
\indent In this paper, we proposed new results in random matrix theory with a specific covariance matrix model fitted to our data model: the \textit{spiked} model. Based on this, we studied the convergence of the traditional estimators of the SINR loss in their full rank and low rank version when the number of secondary data $K\rightarrow\infty$ with a fixed data dimension $m$ and when $m,K\rightarrow\infty$ at the same rate $c=m/K$. We observed that the full rank version is consistent in the two regimes. However, the low rank version is consistent when $K\rightarrow\infty$ with a fixed $m$ but is not consistent when $m,K\rightarrow\infty$ at the same rate $c$. Finally, we applied these results to a jamming application. We first observed that the experimental probability density function of the eigenvalue of the covariance matrix of jamming data is relevant with the probability density function of the \textit{spiked} model. Then, we validated the convergence of the SINR loss in its low rank version and we observed that random matrix theory and more precisely the \textit{spiked} model better evaluate the asymptotic performances of the low rank SINR loss corresponding to the adaptive LR filter, especially when the steering vector parameter is close to the jamming one and contrary to previous works. Moreover, it permits to predict the steering vector parameter value corresponding to the performance break.
\section{Appendix}
\indent The proof is decomposed as follows. We first develop the \textit{structured} QF as a sum of \textit{simple} QFs and base \textit{structured} QF (Subsec. \ref{subsec:A.1}). In a second time, we formulate the base \textit{structured} QF as a complex integral (Subsec. \ref{subsec:A.2}) and split it into several integrals (Subsec. \ref{subsec:A.3}). Then, we determine the deterministic complex integral equivalent of the base \textit{structured} QF (Subsec. \ref{subsec:A.4}) and its formal expression (Subsec. \ref{subsec:A.5}). Finally, we use this result to determine the convergence of the \textit{structured} QF in the large dimensional regime (Subsec. \ref{subsec:A.6}). The regime of convergences in the Appendix, if not precised, is $m,K\rightarrow\infty$ at a fixed ratio $c$.
\subsection{Development of the \textit{structured} QF}\label{subsec:A.1}
\indent Let $\boldsymbol{s}_1$ and $\boldsymbol{s}_2$ be two deterministic complex vectors and $\mathbf{B}$ be a $m\times m$ deterministic complex matrix with uniformly bounded spectral norm for all $m$. In order to obtain the convergence of the \textit{structured} QF $\boldsymbol{s}_1^H\hat{\boldsymbol{\Pi}}_{\mathrm{c}}^{\bot}\mathbf{B}\hat{\boldsymbol{\Pi}}_ {\mathrm{c}}^{\bot} \boldsymbol{s}_2$, one can rewrite, using the notations of Eq.(\ref{eq:SCM_spike}) and the \textit{spiked} model, $\hat{\boldsymbol{\Pi}}_{\mathrm{c}}^{\bot}=\hat{\boldsymbol{\Pi}}_{r+1}=\hat{\mathbf{U}}_{r+1}\hat{\mathbf{U}}_{r+1}^H=\mathbf{I}_m-\sum_{i=1}^r\hat{\boldsymbol{\Pi}}_i$ where $\hat{\mathbf{U}}_{r+1}=[\hat{\boldsymbol{u}}_{r+1},\cdots,\hat{\boldsymbol{u}}_m]$, $\hat{\boldsymbol{\Pi}}_i=\hat{\boldsymbol{u}}_i\hat{\boldsymbol{u}}_i^H$, $\forall i\in[\![1,r]\!]$ and $\hat{\boldsymbol{u}}_i$ are the eigenvectors of the SCM. We recall that $r$ is fixed for all $m$, i.e. $r/m\rightarrow 0^+$. Thus, one can develop the \textit{structured} QF as :\small
\begin{eqnarray}
	\!\!\!\!\!\!\boldsymbol{s}_1^H\hat{\boldsymbol{\Pi}}_{\mathrm{c}}^{\bot}\mathbf{B}\hat{\boldsymbol{\Pi}}_ {\mathrm{c}}^{\bot} \boldsymbol{s}_2\!\!\!\!\!\! &=&\!\!\!\!\!\!\boldsymbol{s}_1^H\left( \mathbf{I}_m-\sum_{i=1}^r\hat{\boldsymbol{\Pi}}_i\right) \mathbf{B}\left(\mathbf{I}_m-\sum_{i=1}^r\hat{\boldsymbol{\Pi}}_i \right)  \boldsymbol{s}_2\\
	&=&\!\!\!\!\!\!\boldsymbol{s}_1^H\mathbf{B}\boldsymbol{s}_2\! -\!\boldsymbol{s}_1^H\sum_{i=1}^r\hat{\boldsymbol{\Pi}}_i\mathbf{B}\boldsymbol{s}_2\! -\!\boldsymbol{s}_1^H\mathbf{B}\sum_{i=1}^r\hat{\boldsymbol{\Pi}}_i\boldsymbol{s}_2 \nonumber\\
	&&+\boldsymbol{s}_1^H\sum_{i=1}^r\hat{\boldsymbol{\Pi}}_i\mathbf{B}\sum_{i=1}^r\hat{\boldsymbol{\Pi}}_i\boldsymbol{s}_2\\
	&=&\!\!\!\!\!\!\boldsymbol{s}_1^H\mathbf{B}\boldsymbol{s}_2\! -\!\sum_{i=1}^r\left(\boldsymbol{s}_1^H\hat{\boldsymbol{\Pi}}_i\mathbf{B}\boldsymbol{s}_2+\boldsymbol{s}_1^H\mathbf{B}\hat{\boldsymbol{\Pi}}_i\boldsymbol{s}_2\right)\nonumber\\
	&&\!\!\!\!\!\! +\!\!\sum_{j_1=1}^r\!\boldsymbol{s}_1^H\hat{\boldsymbol{\Pi}}_{j_1}\mathbf{B}\hat{\boldsymbol{\Pi}}_{j_1}\boldsymbol{s}_2\! +\!\!\!\!\sum_{\underset{j_1\neq j_2}{j_1,j_2=1}}^r\!\!\!\!\boldsymbol{s}_1^H\hat{\boldsymbol{\Pi}}_{j_1}\mathbf{B}\hat{\boldsymbol{\Pi}}_{j_2}\boldsymbol{s}_2\label{eq:FQtot}
\end{eqnarray}\normalsize
\subsection{Formulation of the base \textit{structured} QF as a complex integral}\label{subsec:A.2}
\indent Remarking that Eq.(\ref{eq:FQtot}) is a sum of \textit{simple} QFs and base \textit{structured} QFs, we first focus on the convergence of the base \textit{structured} QF $\hat{\eta}(j_1,j_2)=\boldsymbol{s}_1^H\hat{\boldsymbol{\Pi}}_{j_1}\mathbf{B}\hat{\boldsymbol{\Pi}}_{j_2}\boldsymbol{s}_2$, $\left\lbrace j_1,j_2\right\rbrace \in[\![1,r]\!]^2 $. Let us now formulate the base \textit{structured} QF as a complex integral.\\
\indent \textit{\textbf{Proposition 2:}} Let $\mathbf{B}$ be a $m\times m$ deterministic complex matrix with a uniformly bounded spectral norm for all $m$. Then, under (\textbf{As1}-\textbf{As5}, \textbf{As6.S}) and the \textit{spiked} model, $\forall j_1,j_2\in[\![1,r+1]\!]$, if $\hat{\eta}(j_1,j_2)=\boldsymbol{s}_1^H\hat{\boldsymbol{\Pi}}_{j_1}\mathbf{B}\hat{\boldsymbol{\Pi}}_{j_2}\boldsymbol{s}_2$:\small
\begin{eqnarray}
	\hat{\eta}(j_1,j_2)&=&\dfrac{1}{(2i\pi)^2}\oint_{\mathcal{C}_{j_1}^-}\oint_{\mathcal{C}_{j_2}^-}\boldsymbol{s}_1^H\left(\hat{\mathbf{R}}-z_1\mathbf{I}_m\right)^{-1}\nonumber\\
	&&\times\mathbf{B}\left(\hat{\mathbf{R}}-z_2\mathbf{I}_m\right)^{-1}\boldsymbol{s}_2dz_1dz_2\label{Eq:cauchy_integral}
\end{eqnarray}\normalsize
\begin{flushright} \vspace{-0.3cm}$\blacksquare$ \end{flushright}

\indent \textit{Proof:} If $j_1\neq j_2$, it can be easily shown that $\hat{\eta}(j_1,j_2)$ can be expressed as the following Cauchy integral:\small
\begin{equation}
	A\! =\!\frac{1}{(2i\pi)^2}\!\oint_{\mathcal{C}_{j_1}^-}\!\!\oint_{\mathcal{C}_{j_2}^-}\!\!\!\boldsymbol{s}_1^H\!(\hat{\mathbf{R}}-z_1\mathbf{I}_m)^{-1}\mathbf{B}(\hat{\mathbf{R}}-z_2\mathbf{I}_m)^{-1}\boldsymbol{s}_2dz_1dz_2\label{Eq:IntCurv}
\end{equation}\normalsize
\noindent where $\mathcal{C}_j^-$ in a negatively oriented contour encompassing the eigenvalues of $\hat{\mathbf{R}}$ corresponding to the $j$-th eigenvalue of $\mathbf{R}$ and $z_1$ and $z_2$ are independent variables. Indeed, let $\mathbf{G}(z_k)= (\hat{\mathbf{R}}-z_k\mathbf{I}_m)^{-1}=(\frac{1}{K}\mathbf{X}\mathbf{X}^H-z_k\mathbf{I}_m)^{-1}$ with $k\in\left\lbrace 1,2\right\rbrace $. Thus:\small
\begin{eqnarray}
	\!\!\!\!\!\! A\!\!\!\!\! &=&\!\!\!\!\!\dfrac{1}{(2i\pi)^2}\!\oint_{\mathcal{C}_{j_1}^-}\!\!\oint_{\mathcal{C}_{j_2}^-}\!\!\!\boldsymbol{s}_1^H\mathbf{G}(z_1)\mathbf{B}(\hat{\mathbf{R}}-z_2\mathbf{I}_m)^{-1}\!\boldsymbol{s}_2dz_1dz_2\\
	&=&\!\!\!\!\!\frac{1}{(2i\pi)^2}\!\oint_{\mathcal{C}_{j_1}^-}\!\!\!\oint_{\mathcal{C}_{j_2}^-}\!\!\!\! \boldsymbol{s}_1^H\mathbf{G}(z_1)\mathbf{B}\!\left(\sum_{n=1}^{m}\!\hat{\lambda}_n\hat{\boldsymbol{u}}_n\hat{\boldsymbol{u}}_n^H\! -\! z_2\mathbf{I}_m\!\right)^{-1}\nonumber\\
	&&\qquad\qquad\qquad\qquad\qquad\qquad\qquad\qquad\times\boldsymbol{s}_2dz_2dz_1 \\
	&=&\!\!\!\!\! \frac{1}{(2i\pi)^2}\oint_{\mathcal{C}_{j_1}^-}\oint_{\mathcal{C}_{j_2}^-} \sum_{n=1}^{m}\frac{\boldsymbol{s}_1^H\mathbf{G}(z_1)\mathbf{B}\hat{\boldsymbol{u}}_n\hat{\boldsymbol{u}}_n^H\boldsymbol{s}_2}{\hat{\lambda}_n-z_2}dz_2dz_1 \\
	&=&\!\!\!\!\! \frac{1}{2i\pi}\oint_{\mathcal{C}_{j_1}^-}\sum_{n=1}^{m}\frac{1}{2i\pi}\oint_{\mathcal{C}_{j_2}^-} f_n^{(2)}(z_2)dz_2dz_1
\end{eqnarray}\normalsize\\
\noindent where $f_n^{(2)}(z_2)=\frac{\boldsymbol{s}_1^H\mathbf{G}(z_1)\mathbf{B}\hat{\boldsymbol{u}}_n\hat{\boldsymbol{u}}_n^H\boldsymbol{s}_2}{\hat{\lambda}_n-z_2}$. From the expression $f_n^{(2)}(z_2)$, one observe that $f_n^{(2)}(z_2)$ has a single simple pole $\hat{\lambda}_n$ which is encompassed by $\mathcal{C}_{j_2}^-$ for the indexes $n\in\mathcal{M}_{j_2}$ where $\mathcal{M}_{j_2}$ is the set of indexes corresponding to the $j_2$-th eigenvalue of $\mathbf{R}$. Consequently, from complex analysis:\small
\begin{eqnarray}
	A&=& \frac{1}{2i\pi}\oint_{\mathcal{C}_{j_1}^-}\sum_{n\in\mathcal{M}_{j_2}}\frac{1}{2i\pi}\oint_{\mathcal{C}_{j_2}^-} \frac{\boldsymbol{s}_1^H\mathbf{G}(z_1)\mathbf{B}\hat{\boldsymbol{u}}_{n}\hat{\boldsymbol{u}}_{n}^H\boldsymbol{s}_2}{\hat{\lambda}_n-z_2}dz_2dz_1\nonumber\\
	\\
	&=&\frac{1}{2i\pi}\oint_{\mathcal{C}_{j_1}^-}\sum_{n\in\mathcal{M}_{j_2}}\frac{1}{2i\pi}\oint_{\mathcal{C}_{j_2}^-} f^{(2)}_{n}(z_2)dz_2dz_1 \\
	&=& \frac{1}{2i\pi}\oint_{\mathcal{C}_{j_1}^-}\sum_{n\in\mathcal{M}_{j_2}}\left[ -\mathrm{Res}\left(f^{(2)}_{n}(z_2),\hat{\lambda}_{n}\right)\right] dz_1
\end{eqnarray}\normalsize
where $\mathrm{Res}\left(f^{(2)}_{n}(z_2),\hat{\lambda}_{n}\right)$ is the residue of $f^{(2)}_{n}(z_2)$ at $\hat{\lambda}_n$. Thus, using the residue theorem and residue calculus:\small
\begin{eqnarray}
	\!\!\!\!\!\!\!\! A\!\!\!\!\! &=&\!\!\!\!\!\frac{1}{2i\pi}\oint_{\mathcal{C}_{j_1}^-}\sum_{n\in\mathcal{M}_{j_2}}\left[- \underset{z_2\rightarrow\hat{\lambda}_n}{\mathrm{lim}}(z_2-\hat{\lambda}_n)f^{(2)}_{n}(z_2)\right] dz_1\\
	&=&\!\!\!\!\! \frac{1}{2i\pi}\!\oint_{\mathcal{C}_{j_1}^-}\!\sum_{n\in\mathcal{M}_{j_2}}\!\!\!\underset{z_2\rightarrow\hat{\lambda}_n}{\mathrm{lim}}\!\!\!\left\lbrace  (\hat{\lambda}_n-z_2)\frac{\boldsymbol{s}_1^H\mathbf{G}(z_1)\mathbf{B}\hat{\boldsymbol{u}}_{n}\hat{\boldsymbol{u}}_{n}^H\boldsymbol{s}_2}{\hat{\lambda}_n-z_2}\right\rbrace  dz_1\nonumber \\
	&&\\
	&=&\!\!\!\!\! \frac{1}{2i\pi}\oint_{\mathcal{C}_{j_1}^-}\sum_{n\in\mathcal{M}_{j_2}}\underset{z_2\rightarrow\hat{\lambda}_n}{\mathrm{lim}}\left\lbrace\boldsymbol{s}_1^H\mathbf{G}(z_1)\mathbf{B}\hat{\boldsymbol{u}}_{n}\hat{\boldsymbol{u}}_{n}^H\boldsymbol{s}_2\right\rbrace dz_1
\end{eqnarray}

\begin{eqnarray}
	\!\!\!\!\!\!\!\! A\!\!\!\!\!&=&\!\!\!\!\! \frac{1}{2i\pi}\oint_{\mathcal{C}_{j_1}^-}\boldsymbol{s}_1^H\mathbf{G}(z_1)\mathbf{B}\sum_{n\in\mathcal{M}_{j_2}}\hat{\boldsymbol{u}}_{n}\hat{\boldsymbol{u}}_{n}^H\boldsymbol{s}_2dz_1\\
	&=&\!\!\!\!\! \frac{1}{2i\pi}\oint_{\mathcal{C}_{j_1}^-}\boldsymbol{s}_1^H\left(\hat{\mathbf{R}}-z_1\mathbf{I}_m\right)^{-1}\mathbf{B}\hat{\bm{\Pi}}_{j_2}\boldsymbol{s}_2dz_1\\
	&=&\!\!\!\!\! \frac{1}{2i\pi}\oint_{\mathcal{C}_{j_1}^-}\boldsymbol{s}_1^H\left(\sum_{n=1}^{m}\hat{\lambda}_n\hat{\boldsymbol{u}}_{n}\hat{\boldsymbol{u}}_{n}^H-z_1\mathbf{I}_m\right)^{-1}\mathbf{B}\hat{\bm{\Pi}}_{j_2}\boldsymbol{s}_2dz_1\\
	&=&\!\!\!\!\! \frac{1}{2i\pi}\oint_{\mathcal{C}_{j_1}^-}\sum_{n=1}^{m}\frac{\boldsymbol{s}_1^H\hat{\boldsymbol{u}}_{n}\hat{\boldsymbol{u}}_{n}^H\mathbf{B}\hat{\bm{\Pi}}_{j_2}\boldsymbol{s}_2}{\hat{\lambda}_n-z_1}dz_1\\
	&=&\!\!\!\!\!\sum_{n=1}^{m}\frac{1}{2i\pi}\oint_{\mathcal{C}_{j_1}^-}f^{(1)}_n(z_1)dz_1
\end{eqnarray}\normalsize
where $f_n^{(1)}(z_1)=\frac{\boldsymbol{s}_1^H\hat{\boldsymbol{u}}_{n}\hat{\boldsymbol{u}}_{n}^H\mathbf{B}\hat{\bm{\Pi}}_{j_2}\boldsymbol{s}_2}{\hat{\lambda}_n-z_1}$. Similarly, $f_n^{(1)}(z_1)$ has a single simple pole $\hat{\lambda}_n$ which is encompassed by $\mathcal{C}_{j_1}^-$ for the indexes $n\in\mathcal{M}_{j_1}$. Thus:\small	
\begin{eqnarray}
	A&=& \sum_{n\in\mathcal{M}_{j_1}}\frac{1}{2i\pi}\oint_{\mathcal{C}_{j_1}^-}f^{(1)}_{n}(z_1)dz_1\\
	&=& -\sum_{n\in\mathcal{M}_{j_1}}\mathrm{Res}\left(f^{(1)}_{n}(z_1),\hat{\lambda}_{n}\right)\\
	&=&-\sum_{n\in\mathcal{M}_{j_1}} \underset{z_1\rightarrow\hat{\lambda}_{n}}{\mathrm{lim}}(z_1-\hat{\lambda}_{n})f^{(1)}_{n}(z_1)\\
	&=& \sum_{n\in\mathcal{M}_{j_1}}\underset{z_1\rightarrow\hat{\lambda}_{n}}{\mathrm{lim}}\left\lbrace (\hat{\lambda}_{n}-z_1)\frac{\boldsymbol{s}_1^H\hat{\boldsymbol{u}}_{n}\hat{\boldsymbol{u}}_{n}^H\mathbf{B}\hat{\bm{\Pi}}_{j_2}\boldsymbol{s}_2}{\hat{\lambda}_n-z_1}\right\rbrace \\
	&=& \sum_{n\in\mathcal{M}_{j_1}}\underset{z_1\rightarrow\hat{\lambda}_{n}}{\mathrm{lim}}\left\lbrace \boldsymbol{s}_1^H\hat{\boldsymbol{u}}_{n}\hat{\boldsymbol{u}}_{n}^H\mathbf{B}\hat{\bm{\Pi}}_{j_2}\boldsymbol{s}_2\right\rbrace\\
	&=&\boldsymbol{s}_1^H\sum_{n\in\mathcal{M}_{j_1}}\hat{\boldsymbol{u}}_{n}\hat{\boldsymbol{u}}_{n}^H\mathbf{B}\hat{\bm{\Pi}}_{j_2}\boldsymbol{s}_2=\boldsymbol{s}_1^H\hat{\bm{\Pi}}_{j_1}\mathbf{B}\hat{\bm{\Pi}}_{j_2}\boldsymbol{s}_2
\end{eqnarray}\normalsize
Consequently, $\hat{\eta}(j_1,j_2)=A$ for $j_1\neq j_2$.\\
\indent Then, if $j_1=j_2=j$ and using the same arguments as previously, one has:\small
\begin{eqnarray}
	\boldsymbol{s}_1^H\hat{\bm{\Pi}}_{j}\mathbf{B}\hat{\bm{\Pi}}_{j}\boldsymbol{s}_2=\dfrac{1}{2i\pi}\oint_{\mathcal{C}_{j}^-}\boldsymbol{s}_1^H\sum_{n=1}^m\dfrac{\hat{\boldsymbol{u}}_{n}\hat{\boldsymbol{u}}_{n}^H\mathbf{B}\hat{\boldsymbol{u}}_{n}\hat{\boldsymbol{u}}_{n}^H}{\hat{\lambda}_n-z}\boldsymbol{s}_2dz\label{eq:FQstruct_1int}
\end{eqnarray}\normalsize
However, the remaining of the proof is based on the fact that the resolvent $\mathbf{G}(z)$ of the SCM can be found in the complex integral, which is not the case in the previous equation. Consequently, noticing that:\small
\begin{eqnarray}
	g(\hat{\bm{\Pi}}_{j})=\dfrac{1}{2i\pi}\oint_{\mathcal{C}_{j}^-}\sum_{n=1}^m\dfrac{g(\hat{\bm{\Pi}}_{n})}{\hat{\lambda}_n-z}dz
\end{eqnarray}\normalsize 
where $g(.)$ is a functional, Eq.(\ref{eq:FQstruct_1int}) is equivalent to Eq.(\ref{Eq:cauchy_integral}). As a consequence, $\forall j_1,j_2\in[\![1,r+1]\!]$:\small
\begin{eqnarray}
	\hat{\eta}(j_1,j_2)&=&\dfrac{1}{(2i\pi)^2}\oint_{\mathcal{C}_{j_1}^-}\oint_{\mathcal{C}_{j_2}^-}\boldsymbol{s}_1^H\left(\hat{\mathbf{R}}-z_1\mathbf{I}_m\right)^{-1}\nonumber\\
	&&\times\mathbf{B}\left(\hat{\mathbf{R}}-z_2\mathbf{I}_m\right)^{-1}\boldsymbol{s}_2dz_1dz_2
\end{eqnarray}\normalsize
\subsection{Development of the complex integral}\label{subsec:A.3}
\indent Next, one want to split the previous line integral into several line integrals where some of them will tend to 0. Thus, from~\cite{CoHa13}, with $k\in\left\lbrace 1,2\right\rbrace $, one can write:\small
\begin{eqnarray}
	(\hat{\mathbf{R}}-z_k\mathbf{I}_m)^{-1}\!\!\!\! &=&\!\!\!\! (\mathbf{I}_m+\mathbf{P})^{-1/2}\left[\mathbf{Q}(z_k)-z_k\mathbf{Q}(z_k)\mathbf{U}\right.\nonumber\\
	\!\!\!\! &&\!\!\!\!\left.\times\mathbf{\hat{H}}(z_k)^{-1}\boldsymbol{\Omega}(\mathbf{I}_r+\boldsymbol{\Omega})^{-1}\mathbf{U}^H\mathbf{Q}(z_k) \right] \nonumber\\
	\!\!\!\! &&\!\!\!\! \times(\mathbf{I}_m+\mathbf{P})^{-1/2}\label{Eq:resolvant}
\end{eqnarray}\normalsize
\noindent with\small
\begin{eqnarray}
	\mathbf{Q}(z_k)&=&(\tfrac{1}{K}\mathbf{Y}\mathbf{Y}^H-z_k\mathbf{I}_m)^{-1}\\
	\mathbf{\hat{H}}(z_k)&=&\mathbf{I}_m+z_k\boldsymbol{\Omega}(\mathbf{I}_m+\boldsymbol{\Omega})^{-1}\mathbf{U}^H\mathbf{Q}(z_k)\mathbf{U}
\end{eqnarray}\normalsize
\noindent Then, replacing $(\hat{\mathbf{R}}-z_k\mathbf{I}_m)^{-1}$ by Eq.(\ref{Eq:resolvant}) in Eq.(\ref{Eq:cauchy_integral}) and developing the obtained result, one obtains:\small
\begin{eqnarray}
	\hat{\eta}(j_1,j_2)\!\!\!\! &=& \dfrac{1}{(2i\pi)^2}\oint_{\mathcal{C}_{j_1}^-}\oint_{\mathcal{C}_{j_2}^-}\boldsymbol{s}_1^H \mathbf{E}(z_1)\mathbf{B}\mathbf{E}(z_2)\boldsymbol{s}_2dz_1dz_2\nonumber\\
	&-&\!\!\!\!\dfrac{1}{(2i\pi)^2}\oint_{\mathcal{C}_{j_1}^-}\oint_{\mathcal{C}_{j_2}^-}\left[ \mathbf{\hat{e}}_1^H(z_1) \mathbf{\hat{H}}(z_1)^{-1}\mathbf{\hat{C}}_2(z_1)\mathbf{B}\right]\nonumber\\
	&& \qquad\qquad\qquad\qquad\qquad\quad\times \mathbf{E}(z_2)\boldsymbol{s}_2dz_1dz_2\nonumber\\
	&-&\!\!\!\!\dfrac{1}{(2i\pi)^2}\oint_{\mathcal{C}_{j_1}^-}\oint_{\mathcal{C}_{j_2}^-}\boldsymbol{s}_1^H\mathbf{E}(z_1) \left[\mathbf{B}\mathbf{\hat{C}}_1^H(z_2)\mathbf{\hat{H}}(z_2)^{-1}\right.\nonumber\\
	&&\qquad\qquad\qquad\qquad\qquad\quad\times\left.\mathbf{\hat{e}}_2(z_2) \right]dz_1dz_2\nonumber\\
	&+&\!\!\!\!\dfrac{1}{(2i\pi)^2}\oint_{\mathcal{C}_{j_1}^-}\oint_{\mathcal{C}_{j_2}^-}\mathbf{\hat{e}}_1^H(z_1) \mathbf{\hat{H}}(z_1)^{-1}\mathbf{\hat{C}}_2(z_1)\mathbf{B}\nonumber\\
	&&\qquad\quad\times\mathbf{\hat{C}}_1^H(z_2)\mathbf{\hat{H}}(z_2)^{-1}\mathbf{\hat{e}}_2(z_2)dz_1dz_2\\
	&=& D_1-D_2-D_3+D_4
\end{eqnarray}\normalsize
\noindent with\small
\begin{eqnarray}
	\!\!\!\! \mathbf{E}(z)\!\!\!\! &=&\!\!\!\!(\mathbf{I}_m+\mathbf{P})^{-1/2}\mathbf{Q}(z)(\mathbf{I}_m+\mathbf{P})^{-1/2}\\
    \!\!\!\!\mathbf{\hat{e}}_1^H(z)\!\!\!\! &=&\!\!\!\! \boldsymbol{s}_1^H(\mathbf{I}_m+\mathbf{P})^{-1/2}z\mathbf{Q}(z)\mathbf{U}\\
    \!\!\!\!\mathbf{\hat{C}}_2(z)\!\!\!\! &=&\!\!\!\! \boldsymbol{\Omega}(\mathbf{I}_m+\boldsymbol{\Omega})^{-1}\mathbf{U}^H\mathbf{Q}(z) (\mathbf{I}_m+\mathbf{P})^{-1/2}\\
    \!\!\!\!\mathbf{\hat{C}}_1^H(z)\!\!\!\! &=&\!\!\!\! (\mathbf{I}_m+\mathbf{P})^{-1/2}z\mathbf{Q}(z)\mathbf{U}\\
    \!\!\!\!\mathbf{\hat{e}}_2(z)\!\!\!\! &=&\!\!\!\! \boldsymbol{\Omega}(\mathbf{I}_m+\boldsymbol{\Omega})^{-1}\mathbf{U}^H\mathbf{Q}(z) (\mathbf{I}_m+\mathbf{P})^{-1/2}\boldsymbol{s}_2
\end{eqnarray}\normalsize
\subsection{Determination of the deterministic complex integral equivalent}\label{subsec:A.4}
\indent The convergence of the terms $D_1$ to $D_4$ has now to be studied. Some of them will tend to 0 and the remainder of the terms will tend to a deterministic integral equivalent.\\
\indent \textit{\textbf{Proposition 3:}} Let $\mathbf{B}$ be a $m\times m$ deterministic complex matrix with a uniformly bounded spectral norm for all $m$. Then, under (\textbf{As1}-\textbf{As5}, \textbf{As6.S}) and the \textit{spiked} model, $\forall j_1,j_2\in[\![1,r+1]\!]$, $\hat{\eta}(j_1,j_2)- \eta(j_1,j_2)\overset{\mathrm{a.s.}}{\longrightarrow}0$ with\small
\begin{eqnarray}
	\eta(j_1,j_2)\!\!\!\! &=&\!\!\!\!\dfrac{1}{(2i\pi)^2}\oint_{\gamma_{j_1}^-}\oint_{\gamma_{j_2}^-}\mathbf{e}_1^H(z_1) \mathbf{H}(z_1)^{-1}\mathbf{C}_2(z_1)\nonumber\\
	&&\quad\times \mathbf{B}\mathbf{C}_1^H(z_2)\mathbf{H}(z_2)^{-1}\mathbf{e}_2(z_2)dz_1dz_2\label{eq:eta}
\end{eqnarray}\normalsize
\noindent where $\gamma_{j}^-$ is a deterministic negatively oriented circle only enclosing $\tau_j$ (cf. Eq.(\ref{eq:rho})) and\footnotesize
\begin{eqnarray}
	\mathbf{H}(z)&=&\mathbf{I}_m+z\bar{b}_m(z)\boldsymbol{\Omega}(\mathbf{I}_m+\boldsymbol{\Omega})^{-1}\\
	\mathbf{e}_1^H(z)&=&z\bar{b}_m(z)\boldsymbol{s}_1^H(\mathbf{I}_m+\mathbf{P})^{-1/2}\mathbf{U}\\
	\mathbf{C}_2(z)&=&\bar{b}_m(z)\boldsymbol{\Omega} (\mathbf{I}_m+\boldsymbol{\Omega})^{-1}\mathbf{U}^H (\mathbf{I}_m+\mathbf{P})^{-1/2} \\
	\mathbf{C}_1^H(z)&=&z\bar{b}_m(z) (\mathbf{I}_m+\mathbf{P})^{-1/2}\mathbf{U} \\
	\mathbf{e}_2(z)&=&\bar{b}_m(z) \boldsymbol{\Omega}(\mathbf{I}_m+\boldsymbol{\Omega})^{-1}\mathbf{U}^H (\mathbf{I}_m+\mathbf{P})^{-1/2}\boldsymbol{s}_2
\end{eqnarray}\normalsize
\begin{flushright} \vspace{-0.3cm}$\blacksquare$ \end{flushright}

\indent \textit{Proof:} We first recall that we are interested in the indexes $j_1,j_2\in[\![1,r]\!]$. Then, the function $\mathbf{E}(z)$ in $D_1$, $D_2$ and $D_3$ can be rewritten as:
\begin{eqnarray}
	\mathbf{E}(z)=\left(\hat{\mathbf{R}}-z(\mathbf{I}_m+\mathbf{P})\right)^{-1}=\sum_{n=1}^m\dfrac{\hat{\boldsymbol{u}}_n\hat{\boldsymbol{u}}_n^H}{\hat{\lambda}_n-z(1+\omega_n)}
\end{eqnarray}
Thus, $\mathbf{E}(z_1)$ (resp. $\mathbf{E}(z_2)$) has a single simple pole $\frac{\hat{\lambda}_n}{1+\omega_n}\neq \hat{\lambda}_n$ when $\omega_n\neq 0$, i.e. $n\in[\![1,r]\!]$ ((\textbf{As5, As6.S}) are verified and $\hat{f}(x)\rightarrow f(x)$, with probability one for all large $m,K$ at a fixed ratio $c$). As a consequence, $\forall j_1,j_2\in[\![1,r]\!]$, $\mathcal{C}_{j_1}^-$ (resp. $\mathcal{C}_{j_2}^-$) does not encompass $\mathbf{E}(z_1)$ (resp. $\mathbf{E}(z_2)$). Thus, $D_1=D_2=D_3=0$ and:
\begin{eqnarray}
	\hat{\eta}(j_1,j_2)\!\!\!\! &=&\!\!\!\!\dfrac{1}{(2i\pi)^2}\oint_{\mathcal{C}_{j_1}^-}\oint_{\mathcal{C}_{j_2}^-}\mathbf{\hat{e}}_1^H(z_1) \mathbf{\hat{H}}(z_1)^{-1}\mathbf{\hat{C}}_2(z_1)\mathbf{B}\nonumber\\
	&&\qquad\times\mathbf{\hat{C}}_1^H(z_2)\mathbf{\hat{H}}(z_2)^{-1}\mathbf{\hat{e}}_2(z_2)dz_1dz_2\label{eq:D4}
\end{eqnarray}
\indent We will then determine a deterministic equivalent of Eq.(\ref{eq:D4}), i.e. its convergence in the large dimensional regime from lemma 5 of~\cite{HaLoMeNaVa13}:
\begin{eqnarray}
    \underset{z\in\mathcal{C}}{\mathrm{sup}}\Vert\mathbf{U}^H(\mathbf{Q}(z)-\bar{b}_m(z)\mathbf{I}_m)\mathbf{U}\Vert\underset{\underset{m/K \to c<\infty}{\small{m,K\rightarrow\infty}}}{\overset{\mathrm{a.s.}}{\longrightarrow}}0\label{eq:lemma5}
\end{eqnarray}
\noindent with $\mathcal{C}$ a closed contour of $\C$. Indeed, one can notice that:
\small
\begin{eqnarray}
	\!\!\!\! \mathbf{\hat{H}}(z)\!\!\!\! &=&\!\!\!\!\mathbf{I}_m+z\boldsymbol{\Omega}(\mathbf{I}_m+\boldsymbol{\Omega})^{-1}[\mathbf{U}^H\mathbf{Q}(z)\mathbf{U}]\\
    \!\!\!\!\mathbf{\hat{e}}_1^H(z)\!\!\!\! &=&\!\!\!\! \boldsymbol{s}_1^H(\mathbf{I}_m+\mathbf{P})^{-1/2}z\mathbf{U}[\mathbf{U}^H\mathbf{Q}(z)\mathbf{U}]\\
    \!\!\!\!\mathbf{\hat{C}}_2(z)\!\!\!\! &=&\!\!\!\! \boldsymbol{\Omega}(\mathbf{I}_m+\boldsymbol{\Omega})^{-1}[\mathbf{U}^H\mathbf{Q}(z)\mathbf{U}]\mathbf{U}^H (\mathbf{I}_m+\mathbf{P})^{-1/2}\\
    \!\!\!\!\mathbf{\hat{C}}_1^H(z)\!\!\!\! &=&\!\!\!\! (\mathbf{I}_m+\mathbf{P})^{-1/2}z\mathbf{U}[\mathbf{U}^H\mathbf{Q}(z)\mathbf{U}]\\
    \!\!\!\!\mathbf{\hat{e}}_2(z)\!\!\!\! &=&\!\!\!\! \boldsymbol{\Omega}(\mathbf{I}_m+\boldsymbol{\Omega})^{-1}[\mathbf{U}^H\mathbf{Q}(z) \mathbf{U}]\mathbf{U}^H(\mathbf{I}_m+\mathbf{P})^{-1/2}\boldsymbol{s}_2
\end{eqnarray}\normalsize
\noindent Thus, from Eq.(\ref{eq:lemma5}), one obtains:
\footnotesize
\begin{eqnarray}
	\!\!\!\!\!\!\!\!\mathbf{\hat{H}}(z)\!\!\!\!\!\!\!\!\!\!\!\!  &\underset{\underset{m/K \to c<\infty}{\small{m,K\rightarrow\infty}}}{\overset{\mathrm{a.s.}}{\longrightarrow}}&\!\!\!\!\!\!\!\!\!\!\!\! \mathbf{H}(z)=\mathbf{I}_m+z\bar{b}_m(z)\boldsymbol{\Omega}(\mathbf{I}_m+\boldsymbol{\Omega})^{-1}\\
	\!\!\!\!\!\!\!\!\mathbf{\hat{e}}_1^H(z)\!\!\!\!\!\!\!\!\!\!\!\!  &\underset{\underset{m/K \to c<\infty}{\small{m,K\rightarrow\infty}}}{\overset{\mathrm{a.s.}}{\longrightarrow}}&\!\!\!\!\!\!\!\!\!\!\!\!\mathbf{e}_1^H(z)=z\bar{b}_m(z)\boldsymbol{s}_1^H(\mathbf{I}_m+\mathbf{P})^{-1/2}\mathbf{U}\\
	\!\!\!\!\!\!\!\!\mathbf{\hat{C}}_2(z)\!\!\!\!\!\!\!\!\!\!\!\!  &\underset{\underset{m/K \to c<\infty}{\small{m,K\rightarrow\infty}}}{\overset{\mathrm{a.s.}}{\longrightarrow}}&\!\!\!\!\!\!\!\!\!\!\!\! \mathbf{C}_2(z)=\bar{b}_m(z)\boldsymbol{\Omega} (\mathbf{I}_m+\boldsymbol{\Omega})^{-1}\mathbf{U}^H (\mathbf{I}_m+\mathbf{P})^{-1/2} \\
	\!\!\!\!\!\!\!\!\mathbf{\hat{C}}_1^H(z)\!\!\!\!\!\!\!\!\!\!\!\!  &\underset{\underset{m/K \to c<\infty}{\small{m,K\rightarrow\infty}}}{\overset{\mathrm{a.s.}}{\longrightarrow}}&\!\!\!\!\!\!\!\!\!\!\!\!  \mathbf{C}_1^H(z)=z\bar{b}_m(z) (\mathbf{I}_m+\mathbf{P})^{-1/2}\mathbf{U} \\
	\!\!\!\!\!\!\!\!\mathbf{\hat{e}}_2(z)\!\!\!\!\!\!\!\!\!\!\!\!  &\underset{\underset{m/K \to c<\infty}{\small{m,K\rightarrow\infty}}}{\overset{\mathrm{a.s.}}{\longrightarrow}}&\!\!\!\!\!\!\!\!\!\!\!\!  \mathbf{e}_2(z)=\bar{b}_m(z) \boldsymbol{\Omega}(\mathbf{I}_m+\boldsymbol{\Omega})^{-1}\mathbf{U}^H (\mathbf{I}_m+\mathbf{P})^{-1/2}\boldsymbol{s}_2
\end{eqnarray}\normalsize
\noindent As a result, $\hat{\eta}(j_1,j_2)- \eta(j_1,j_2)\overset{\mathrm{a.s.}}{\longrightarrow}0$ with\small
\begin{eqnarray}
	\eta(j_1,j_2)\!\!\!\! &=&\!\!\!\!\dfrac{1}{(2i\pi)^2}\oint_{\gamma_{j_1}^-}\oint_{\gamma_{j_2}^-}\mathbf{e}_1^H(z_1) \mathbf{H}(z_1)^{-1}\mathbf{C}_2(z_1)\nonumber\\
	&&\quad\times \mathbf{B}\mathbf{C}_1^H(z_2)\mathbf{H}(z_2)^{-1}\mathbf{e}_2(z_2)dz_1dz_2\label{eq:eta}
\end{eqnarray}\normalsize
\noindent where $\gamma_{j}^-$ is a deterministic negatively oriented circle only enclosing $\tau_j$ (cf. Eq.(\ref{eq:rho})). 
\subsection{Determination of the expression of the deterministic equivalent}\label{subsec:A.5}
\indent Let us now find the expression of the deterministic equivalent $\eta(j_1,j_2)$ as a function of the eigenvalues and eigenvectors of the covariance matrix $\mathbf{R}$.\\
\indent \textit{\textbf{Proposition 4:}} Let $\mathbf{B}$ be a $m\times m$ deterministic complex matrix with a uniformly bounded spectral norm for all $m$. Then, under (\textbf{As1}-\textbf{As5}, \textbf{As6.S}) and the \textit{spiked} model, \small
\begin{eqnarray}
	\eta(j_1,j_2)=\chi_{j_1}\chi_{j_2}\boldsymbol{s}_1^H\bm{\Pi}_{j_1}\mathbf{B}\bm{\Pi}_{j_2} \boldsymbol{s}_2 \label{eq:FQ3}
\end{eqnarray}\normalsize
with $\chi_j=\frac{1-c\omega_j^{-2}}{1+c\omega_j^{-1}}$ and $\left\lbrace j_1,j_2\right\rbrace\in[\![1,r]\!]^2$. 
\begin{flushright} \vspace{-0.3cm}$\blacksquare$ \end{flushright}

\indent \textit{Proof:} We first rewrite Eq.(\ref{eq:eta}) as: \small
\begin{eqnarray}
	\eta(j_1,j_2) =\dfrac{1}{2i\pi}\oint_{\gamma_{j_2}^-}\mathbf{g}\times \mathbf{B}\mathbf{C}_1^H(z_2)\mathbf{H}(z_2)^{-1}\mathbf{e}_2(z_2)dz_2\label{eq:eta2}
\end{eqnarray}\normalsize
with\small
\begin{eqnarray}
	\mathbf{g}=\dfrac{1}{2i\pi}\oint_{\gamma_{j_1}^-}\mathbf{e}_1^H(z_1) \mathbf{H}(z_1)^{-1}\mathbf{C}_2(z_1)dz_1
\end{eqnarray}\normalsize
\noindent in order to determine $\mathbf{g}$ in a first time.\\
\noindent We recall that, in our case, $\omega_1>\cdots>\omega_r>\omega_{r+1}=0$. After an eigendecomposition of $\mathbf{e}_1^H(z_1)$ and $\mathbf{C}_2(z_1)$ and, noticing from~\cite{CoHa13} that:\small
\begin{eqnarray}
	\!\! \mathbf{H}(z)^{-1}\!\!\!\!\!\! &=&\!\!\!\!\!\!\mathrm{diag}\left(\tfrac{1}{1+z\bar{b}_m(z)\frac{\omega_1}{1+\omega_1}},\cdots,\tfrac{1}{1+z\bar{b}_m(z)\frac{\omega_{r+1}}{1+\omega_{r+1}}}\right)\\
	\!\!\!\!\!\! &=&\!\!\!\!\!\!\sum_{l=1}^{r+1}\dfrac{1}{1+z\bar{b}_m(z)\frac{\omega_l}{1+\omega_l}}\boldsymbol{\mathcal{I}}_l
	\label{eq:H}
\end{eqnarray}\normalsize
\noindent with\small
\begin{eqnarray}
	\boldsymbol{\mathcal{I}}_l=\begin{bmatrix}
		\boldsymbol{\Nul}_{\mathcal{K}_1+\ldots +\mathcal{K}_{l-1}} && \\
		&\mathbf{I}_{\mathcal{K}_l}& \\
		&& \boldsymbol{\Nul}_{\mathcal{K}_{l+1}+\ldots +\mathcal{K}_{r+1}}
	\end{bmatrix}\in \C^{m\times m}
\end{eqnarray}\normalsize
\noindent one obtains:\small
\begin{eqnarray}
	\mathbf{e}_1^H(z_1) \mathbf{H}(z_1)^{-1}\mathbf{C}_2(z_1)= \boldsymbol{s}_1^H\sum_{l=1}^{r+1}\tfrac{\omega_l \bm{\Pi}_l}{(1+\omega_l)^2}\tfrac{z_1\bar{b}_m^2(z_1)}{1+z_1\bar{b}_m(z_1)\frac{\omega_l}{1+\omega_l}}
\end{eqnarray}\normalsize
Thus,\small
\begin{eqnarray}
	\mathbf{g}&=&\dfrac{1}{2i\pi}\oint_{\gamma_{j_1}^-}\boldsymbol{s}_1^H\sum_{l=1}^{r+1}\tfrac{\omega_l \bm{\Pi}_l}{(1+\omega_l)^2}\tfrac{z_1\bar{b}_m^2(z_1)}{1+z_1\bar{b}_m(z_1)\frac{\omega_l}{1+\omega_l}}dz_1\\
	&=&\sum_{l=1}^{r+1}\tfrac{\omega_l }{(1+\omega_l)^2}\boldsymbol{s}_1^H\bm{\Pi}_l\dfrac{1}{2i\pi}\oint_{\gamma_{j_1}^-}\tfrac{z_1\bar{b}_m^2(z_1)}{1+z_1\bar{b}_m(z_1)\frac{\omega_l}{1+\omega_l}}dz_1\\
	&=&\sum_{l=1}^{r+1}\tfrac{1}{1+\omega_l}\boldsymbol{s}_1^H\bm{\Pi}_l\dfrac{1}{2i\pi}\oint_{\gamma_{j_1}^-}\tfrac{z_1\bar{b}_m^2(z_1)}{\frac{1+\omega_l}{\omega_l}+z_1\bar{b}_m(z_1)}dz_1
\end{eqnarray}\normalsize
From~\cite{CoHa13}, $\frac{1+\omega_l}{\omega_l}+z_1\bar{b}_m(z_1)=0$ only for $z_1=\tau_{j_1}$ and $z_1\bar{b}_m^2(z_1)\vert_{z_1=\tau_{j_1}}\neq 0$, $j_1\in[\![1,r]\!]$. Hence, $\tfrac{z_1\bar{b}_m^2(z_1)}{\frac{1+\omega_l}{\omega_l}+z_1\bar{b}_m(z_1)}$ has a single simple pole at $\tau_{j_1}$, $j_1\in[\![1,r]\!]$. As a consequence, with $h(z)=z\bar{b}_m(z)$\small
\begin{eqnarray}
	\mathbf{g}&=&\tfrac{1}{1+\omega_{j_1}}\boldsymbol{s}_1^H\bm{\Pi}_{j_1}\dfrac{1}{2i\pi}\oint_{\gamma_{j_1}^-}\tfrac{z_1\bar{b}_m^2(z_1)}{\frac{1+\omega_{j_1}}{\omega_{j_1}}+z_1\bar{b}_m(z_1)}dz_1\\
	&=&\tfrac{1}{1+\omega_{j_1}}\boldsymbol{s}_1^H\bm{\Pi}_{j_1}\left[-\mathrm{Res}\left(\dfrac{h(z_1)\bar{b}_m(z_1)}{\frac{1+\omega_{j_1}}{\omega_{j_1}}+h(z_1)},\tau_{j_1}\right)  \right] \\
	&=&\tfrac{1}{1+\omega_{j_1}}\boldsymbol{s}_1^H\bm{\Pi}_{j_1}\left[-\dfrac{h(\tau_{j_1})\bar{b}_m(\tau_{j_1})}{\left.\left( \frac{1+\omega_{j_1}}{\omega_{j_1}}+h(z_1)\right)'\right\vert_{z_1=\tau_{j_1}}}\right]\\
	&=&\tfrac{1}{1+\omega_{j_1}}\boldsymbol{s}_1^H\bm{\Pi}_{j_1}\left[-\dfrac{h(\tau_{j_1})\bar{b}_m(\tau_{j_1})}{h'(\tau_{j_1})}\right]
\end{eqnarray}\normalsize
from residue calculus and where $(.)'=\frac{d(.)}{dz}$. Then, observing that $\frac{1}{1+\omega_{j_1}}=\frac{1+h(\tau_{j_1})}{h(\tau_{j_1})}$ as $\frac{1+\omega_{j_1}}{\omega_{j_1}}+h(\tau_{j_1})=0$, one finally has\small
\begin{eqnarray}
	\mathbf{g}&=&-\boldsymbol{s}_1^H\bm{\Pi}_{j_1}\frac{1+h(\tau_{j_1})}{h(\tau_{j_1})}\dfrac{h(\tau_{j_1})\bar{b}_m(\tau_{j_1})}{h'(\tau_{j_1})}\\
	&=&-\boldsymbol{s}_1^H\bm{\Pi}_{j_1}\dfrac{(1+h(\tau_{j_1}))\bar{b}_m(\tau_{j_1})}{h'(\tau_{j_1})}
\end{eqnarray}\normalsize
Then, similarly, we write Eq.(\ref{eq:eta2}) as:\small
\begin{eqnarray}
	\eta(j_1,j_2) =\mathbf{g} \mathbf{B}\times\mathbf{\tilde{g}}
\end{eqnarray}\normalsize
with\small
\begin{eqnarray}
	\mathbf{\tilde{g}}=\dfrac{1}{2i\pi}\oint_{\gamma_{j_2}^-}\mathbf{C}_1^H(z_2)\mathbf{H}(z_2)^{-1}\mathbf{e}_2(z_2)dz_2
\end{eqnarray}\normalsize
\noindent Thus, similarly as with $\mathbf{g}$, one deduces that:
\small
\begin{eqnarray}
	\mathbf{\tilde{g}}=-\dfrac{(1+h(\tau_{j_2}))\bar{b}_m(\tau_{j_2})}{h'(\tau_{j_2})}\bm{\Pi}_{j_2}\boldsymbol{s}_2
\end{eqnarray}\normalsize
As a result:\small
\begin{eqnarray}
	\eta(j_1,j_2)=\xi(\tau_{j_1})\xi(\tau_{j_2})\boldsymbol{s}_1^H\bm{\Pi}_{j_1}\mathbf{B}\bm{\Pi}_{j_2}\boldsymbol{s}_2\label{Eq:3}
\end{eqnarray}\normalsize
\noindent with\small
\begin{eqnarray}
	\xi(\tau_{j})=\dfrac{(1+h(\tau_j))\bar{b}_m(\tau_j)}{h'(\tau_j)}
\end{eqnarray}\normalsize
\indent Finally, the last step consists in expressing $\xi(\tau_{j})$ as a function of $\omega_j$. Using Corollary 2 from~\cite{CoHa13}, one expresses $\xi(\tau_{j})$ as:\small
\begin{eqnarray}
	\xi(\tau_{j})=\chi_j=\frac{1-c\omega_j^{-2}}{1+c\omega_j^{-1}}
\end{eqnarray}\normalsize
As a consequence,\small
\begin{eqnarray}
	\hat{\eta}(j_1,j_2)\underset{\underset{m/K \to c<\infty}{\small{m,K\rightarrow\infty}}}{\overset{\mathrm{a.s.}}{\longrightarrow}}\eta(j_1,j_2)=\chi_{j_1}\chi_{j_2}\boldsymbol{s}_1^H\bm{\Pi}_{j_1}\mathbf{B}\bm{\Pi}_{j_2} \boldsymbol{s}_2 \label{eq:FQ3}
\end{eqnarray}\normalsize
\noindent with $\left\lbrace j_1,j_2\right\rbrace\in[\![1,r]\!]^2$. 
\subsection{Convergence of the \textit{structured} QF}\label{subsec:A.6}
\indent From the development of the \textit{structured} QF, we recall that the convergences of the \textit{simple} QFs $\boldsymbol{s}_1^H\hat{\boldsymbol{\Pi}}_i\mathbf{B}\boldsymbol{s}_2$ and $\boldsymbol{s}_1^H\mathbf{B}\hat{\boldsymbol{\Pi}}_i\boldsymbol{s}_2$, $\forall i\in[\![1,r]\!]$ can be easily determined from~\cite{CoHa13}:\small
\begin{eqnarray}
	\boldsymbol{s}_1^H\hat{\boldsymbol{\Pi}}_i\mathbf{B}\boldsymbol{s}_2\underset{\underset{m/K\rightarrow c<\infty}{m,K\rightarrow\infty}}{\overset{\text{a.s}}{\longrightarrow}}\chi_i\boldsymbol{s}_1^H\boldsymbol{\Pi}_i\mathbf{B}\boldsymbol{s}_2\label{eq:FQ1}\\
	\boldsymbol{s}_1^H\mathbf{B}\hat{\boldsymbol{\Pi}}_i\boldsymbol{s}_2\underset{\underset{m/K\rightarrow c<\infty}{m,K\rightarrow\infty}}{\overset{\text{a.s}}{\longrightarrow}}\chi_i\boldsymbol{s}_1^H\mathbf{B}\boldsymbol{\Pi}_i\boldsymbol{s}_2 \label{eq:FQ2}
\end{eqnarray}\normalsize
\noindent where $\chi_i$ is defined as in Section III.C.\\
\indent Then, also using Eq.(\ref{eq:FQ3}) in Eq.(\ref{eq:FQtot}), one obtains:\small
\begin{eqnarray}
	\!\!\!\!\!\!\boldsymbol{s}_1^H\hat{\boldsymbol{\Pi}}_{\mathrm{c}}^{\bot}\mathbf{B}\hat{\boldsymbol{\Pi}}_ {\mathrm{c}}^{\bot} \boldsymbol{s}_2\!\!\!\!\!\! &\underset{\underset{m/K \to c<\infty}{\small{m,K\rightarrow\infty}}}{\overset{\mathrm{a.s.}}{\longrightarrow}}&\!\!\!\!\!\!\boldsymbol{s}_1^H\mathbf{B}\boldsymbol{s}_2\nonumber\\
	&&\!\!\!\!\!\!\!\!\!\!\!\!\!\!\!\!\!\! -\!\sum_{i=1}^r\left(\boldsymbol{s}_1^H\chi_i\boldsymbol{\Pi}_i\mathbf{B}\boldsymbol{s}_2+\boldsymbol{s}_1^H\mathbf{B}\chi_i\boldsymbol{\Pi}_i\boldsymbol{s}_2\right)\nonumber\\
	&&\!\!\!\!\!\!\!\!\!\!\!\!\!\!\!\!\!\! +\!\!\sum_{j_1=1}^r\!\boldsymbol{s}_1^H\chi_{j_1}\boldsymbol{\Pi}_{j_1}\mathbf{B}\chi_{j_1}\boldsymbol{\Pi}_{j_1}\boldsymbol{s}_2\nonumber\\
	&&\!\!\!\!\!\!\!\!\!\!\!\!\!\!\!\!\!\! +\!\!\!\!\sum_{\underset{j_1\neq j_2}{j_1,j_2=1}}^r\!\!\!\!\boldsymbol{s}_1^H\chi_{j_1}\boldsymbol{\Pi}_{j_1}\mathbf{B}\chi_{j_2}\boldsymbol{\Pi}_{j_2}\boldsymbol{s}_2
\end{eqnarray}\normalsize
or equivalently\small
\begin{eqnarray}
	&&\!\!\!\!\!\!\!\!\!\!\!\!\!\!\!\!\!\!\!\!\boldsymbol{s}_1^H\hat{\boldsymbol{\Pi}}_{\mathrm{c}}^{\bot}\mathbf{B}\hat{\boldsymbol{\Pi}}_ {\mathrm{c}}^{\bot} \boldsymbol{s}_2\underset{\underset{m/K \to c<\infty}{\small{m,K\rightarrow\infty}}}{\overset{\mathrm{a.s.}}{\longrightarrow}}\nonumber\\
	&&\qquad\boldsymbol{s}_1^H\left[ \mathbf{I}_m-\sum_{i=1}^r\chi_i\boldsymbol{\Pi}_i\right] \mathbf{B}\left[\mathbf{I}_m-\sum_{i=1}^r\chi_i\boldsymbol{\Pi}_i \right]  \boldsymbol{s}_2
\end{eqnarray}\normalsize\small
\begin{eqnarray}
	\boldsymbol{s}_1^H\hat{\boldsymbol{\Pi}}_{\mathrm{c}}^{\bot}\mathbf{B}\hat{\boldsymbol{\Pi}}_ {\mathrm{c}}^{\bot} \boldsymbol{s}_2\underset{\underset{m/K \to c<\infty}{\small{m,K\rightarrow\infty}}}{\overset{\mathrm{a.s.}}{\longrightarrow}}\boldsymbol{s}_1^H\bar{\boldsymbol{\Pi}}_{\mathrm{c,S}}^{\bot} \mathbf{B}\bar{\boldsymbol{\Pi}}_{\mathrm{c,S}}^{\bot}\boldsymbol{s}_2
\end{eqnarray}\normalsize
\noindent with $\bar{\boldsymbol{\Pi}}_{\mathrm{c,S}}^{\bot}=\sum_{i=1}^{m} \psi_i\boldsymbol{u}_i\boldsymbol{u}_i^H$ and\small
\begin{eqnarray}
	\psi_i=\begin{cases}1,\;\,\qquad\quad \mathrm{if}\;i>r\\
		1-\chi_i,\;\quad \mathrm{if}\;i\leqslant r
	\end{cases}
\end{eqnarray}\normalsize

\footnotesize
\bibliographystyle{IEEEtran} 
\bibliography{Biblio} 

\end{document}